\documentclass[prb, preprint, twocolumn, amsmath, amssymb, reprint,
superscriptaddress,
]{revtex4-1}

\usepackage{graphicx,verbatim}
\usepackage{dcolumn}
\usepackage{bm}
\usepackage{sidecap}
\usepackage{braket}
\usepackage{color}

\begin{document}

\title{Vibrationally dependent electron-electron interactions in resonant electron transport through single-molecule junctions}

\author{A.\ Erpenbeck}
\affiliation{Institut f\"ur Theoretische Physik und Interdisziplin\"ares Zentrum f\"ur Molekulare Materialien, Friedrich-Alexander-Universit\"at Erlangen-N\"urnberg, Staudtstr.\, 7/B2, D-91058 Erlangen, Germany}
\author{R.\ H\"artle}
\affiliation{Institut f\"ur theoretische Physik, Georg-August-Universit\"at G\"ottingen, Friedrich-Hund-Platz 1, D-37077 G\"ottingen, Germany}
\author{M.\ Bockstedte}
\affiliation{Institut f\"ur Theoretische Physik und Interdisziplin\"ares Zentrum f\"ur Molekulare Materialien, Friedrich-Alexander-Universit\"at Erlangen-N\"urnberg, Staudtstr.\, 7/B2, D-91058 Erlangen, Germany}
\author{M.\ Thoss}
\affiliation{Institut f\"ur Theoretische Physik und Interdisziplin\"ares Zentrum f\"ur Molekulare Materialien, Friedrich-Alexander-Universit\"at Erlangen-N\"urnberg, Staudtstr.\, 7/B2, D-91058 Erlangen, Germany}

\date{\today}

\begin{abstract}
	We investigate the role of electronic-vibrational coupling in resonant electron transport through 
	single-molecule junctions, taking into account that the corresponding coupling strengths 
	may depend on the charge and excitation state of the molecular bridge. 
	In the presence of multiple electronic states, this requires to extend the commonly used model and include vibrationally dependent electron-electron interaction. We use Born-Markov 
	master equation methods and consider selected models to exemplify the effect of the additional interaction on the transport characteristics of 
	a single-molecule junction. In particular, we show that it has a significant influence on local cooling and heating mechanisms, may result in negative differential resistance, and cause  
	pronounced asymmetries in the conductance map of a single-molecule junction.

\end{abstract}


\maketitle

\section{Introduction}

	Electron transport through a single-molecule junction constitutes a complex many-body problem. Electronic and vibrational degrees of freedom of a molecular conductor are often strongly correlated, in particular in nonequilibrium states at higher bias voltages \cite{Galperin07, WuNazinHo04, LeRoy, Natelson04, Pasupathy05, Sapmaz2006, Thijssen06, Parks07, Boehler07, Leon2008, Huettel2009, Tao2010, Ballmann2010, Jewell2010, Osorio2010, Koch2005, Hartle09, Romano10, Secker2010, Hartle2011b, Ballmann2013b, Hartle2013}. 
	The investigation of coupling between electronic and nuclear degrees of freedom in nanostructures under nonequilibrium conditions has been of great interest recently and revealed a wealth of physical phenomena such as rectification \cite{Braig2003, WuNazinHo04, Zazunov2006, Haertle11, Erpenbeck2015}, vibrationally induced decoherence \cite{Haertle11_3, Ballmann2012, Haertle13, Erpenbeck2015} negative differential resistance \cite{Schoeller01, Koch05b, Zazunov06, Leijnse09, Hartle2010b, Haertle11, Erpenbeck2015}, and signatures of multistability \cite{Galperin2005, Thoss2012, Albrecht2013, Wilner2014}. 

 	Most theoretical studies of vibrationally coupled electron transport in molecular junctions have employed a simplified model of a molecule with linear coupling of the electronic degrees of freedom to vibrational modes described in the harmonic approximation \cite{Cederbaum74, Galperin07, Benesch08, Han2010, Haertle11}
	More realistic models include charge-dependent vibrational frequencies 
	\cite{Wegewijs05, Koch05b, White2012, Kaasbjerg2013} or anharmonic nuclear potentials \cite{Cizek05, Koch05b, Donarini2006, Huebner2007, Elste2008, Pshenichnyuk2010, Brueggemann2012, Simine2014}. 
	The description of systems with multiple electronic states often requires to include the Coulomb interaction between electrons. The corresponding interaction strengths depend on the specific electronic configuration and on the nuclear geometry. This results in vibrationally dependent electron-electron interactions. The study of this effect and its manifestation in transport characteristics of molecular junctions is the main subject of this article.
	The importance of vibrationally dependent electron-electron interactions has already been realized, for example, in the context of the Hubbard model approach to electron-electron interaction in molecules \cite{Schmalz2011} as well as in studies of dissociation in colloidal quantum dot systems \cite{Pozner2014}. 
		
	The outline of this article is as follows: The theoretical methodology is introduced in Sec.\ \ref{sec:Theory}, including the model Hamiltonian of the molecular junctions,  the master equation approach used to describe charge transport, and a discussion of the relevance of vibrationally dependent electron-electron interactions. 
	In Sec.\ \ref{sec:Results}, we analyze the effects and manifestation of vibrationally dependent electron-electron interactions in transport characteristics. To this end, we consider selected model systems, including  scenarios with symmetric and asymmetric molecule-lead coupling.

\section{Theoretical methodology}\label{sec:Theory}

\subsection{Model Hamiltonian}\label{sec:Hamiltonian}

	The Hamiltonian of the molecular bridge in atomic units is given by
		\begin{subequations}\label{eq:Hamiltonian_1}
		\begin{eqnarray}
		H_{\text{S}} &=& H_{\text{nuc}} + H_{\text{el}}, \\
		H_{\text{nuc}} &=& - \sum_{a} \frac{1}{2M_{a}} \Delta_{a} +  V_{\text{nuc}}(\textbf{R}),  \\
		V_{\text{nuc}}(\textbf{R}) &=& \sum_{a<b} \frac{Z_{a}Z_{b}}{\left\vert \textbf{R}_{a} - \textbf{R}_{b} \right\vert}, \\
		H_{\text{el}}(\textbf{R}) &=& - \sum_{i} \frac{\Delta_{i}}{2}  + \sum_{i<j} \frac{1}{\left\vert \textbf{r}_{i} - \textbf{r}_{j} \right\vert} \nonumber \\ &&
					 - \sum_{i a}\frac{Z_{a}}{\left\vert \textbf{R}_{a} - \textbf{r}_{i} \right\vert},
		\end{eqnarray}
		\end{subequations}
	where $H_{\text{nuc}}$ describes the nuclear and $H_{\text{el}}(\textbf{R})$ the electronic degrees of freedom of the molecule. The nuclear part of the Hamiltonian, $H_{\text{nuc}}$,  includes the kinetic energy of the nuclei and the Coulomb repulsion, $ V_{\text{nuc}}(\textbf{R})$.  Similarly, $H_{\text{el}}(\textbf{R})$ includes the kinetic energy of the electrons, the Coulomb repulsion, and the Coulomb attraction between the electrons and the nuclei. Thereby, the vector $\textbf{R}$ summarizes the coordinates $\textbf{R}_{a}$ of the nuclei and $\textbf{r}_{i}$ denotes the coordinates of the $i$th electron.  The charge and the mass of the nuclei are given by $Z_{a}$ and $M_{a}$, respectively. 

	We follow the scheme outlined in \cite{Cederbaum74, Cederbaum76, Benesch06, Kondov07, Benesch08, Benesch2009, HartlePhD} to derive an approximate representation of the Hamiltonian in second quantization. Thereby we focus on the molecular electronic states, thus working in a restricted subspace only incorporating a subset of all possible states. Accordingly, the result will be a restricted model Hamiltonian, containing effective interaction terms. Those effective interaction terms not only account for the bare interaction, but also for the influence of the states beyond the restricted subspace under consideration. As a reference state we use the electronic ground state of the uncharged molecule
		\begin{eqnarray}
		H_{\text{el}}(\textbf{R}) \ket{\Psi_{\text{ref}}(\mathbf{r;R})}&=& E_{\text{el, ref}}(\textbf{R}) \ket{\Psi_{\text{ref}}(\mathbf{r;R})},
		\end{eqnarray}
	which depends parametrically on the nuclear coordinates $\textbf{R}$. Employing an effective single particle description (e.g.\ Hartree-Fock or density functional theory), the electronic ground state is given by a single Slater determinant,
		\begin{eqnarray}
		\ket{\Psi_{\text{ref}}(\mathbf{r;R})} &=& \prod_{m \in \lbrace \text{occ.} \rbrace} d_m^\dagger(\mathbf{R}) \ket{0}.
		\end{eqnarray}
	This determinant involves single-particle states (orbitals) $\ket{\varphi_m(\mathbf{r}_i;\mathbf{R})}$ which are occupied in the reference state. Using the creation and annihilation operators $d_m^{(\dagger)}(\mathbf{R})$ corresponding to $\ket{\varphi_m(\mathbf{r}_i;\mathbf{R})}$, the electronic part of the Hamiltonian can be rewritten as \cite{Cederbaum74, Cederbaum76, HartlePhD}
		\begin{eqnarray} \label{eq:H_el_second}
		H_{\text{el}}(\textbf{R})  &=& E_{\text{el, ref}}(\textbf{R}) + \sum_{m} \epsilon_{m}(\textbf{R}) (d_{m}^{\dagger}d_{m}-\delta_{m}) \\
		&& +  \sum_{mn} U_{mn}(\textbf{R}) (d_{m}^{\dagger}d_{m}-\delta_{m})(d_{n}^{\dagger}d_{n}-\delta_{n}),  \nonumber
		\end{eqnarray}
	where $\epsilon_{m}(\mathbf{R})$ denote the orbital energies. 
	The parameters $\delta_m$ distinguish between single-particle states that are occupied ($\delta_{m}=1$) or unoccupied ($\delta_{m}=0$) in the reference state $\ket{\Psi_{\text{ref}}(\mathbf{r;R})}$. 
	The energy of a molecular state that differs in the population of two electronic orbitals $m$ and $n$ from the reference state is not only modified by the energy differences $\epsilon_{m}(\mathbf{R})$ and $\epsilon_{n}(\mathbf{R})$, but also by the Coulomb interaction between the electrons $U_{mn}(\mathbf{R})$. 
	In principle, there is also the necessity for interaction terms describing the change in energy for molecular sates that differ in the population of three or more electrons from the ground state. They are, however, beyond the scope of our present considerations. Moreover, in the derivation of Eq.\ (\ref{eq:H_el_second}), we have applied the adiabatic approximation, that is neglecting the coupling between different electronic states due to electronic-vibrational or electron-electron interactions \cite{HartlePhD, Erpenbeck2015}. Such effects are considered, for example, in Refs.\ \onlinecite{Reckermann2008, Frederiksen2008, Schultz2008, Repp2009, Han2010, Erpenbeck2015}. For simplicity, we furthermore, do not consider the spin explicitely.  

	To characterize the nuclear (vibrational) degrees of freedom, we employ the normal modes of the reference state $\ket{\Psi_{\text{ref}}(\mathbf{r;R})}$. The corresponding potential energy surface (PES) is given by $ V_{\text{nuc}}(\textbf{R})+E_{\text{el, ref}}(\textbf{R})$. Using the harmonic approximation for the PES and dropping an irrelevant constant, the vibrational part of $H_{\text{S}}$ can be rewritten as  
		\begin{eqnarray}
		- \sum_{a} \frac{1}{2M_{a}} \Delta_{a} +  V_{\text{nuc}}(\textbf{R}) +  E_{\text{el, ref}}(\textbf{R})	\approx \sum_{\alpha} \Omega_{\alpha} a_{\alpha}^{\dagger}a_{\alpha},   \nonumber \\
		\end{eqnarray}
	where the ladder operators $a_{\alpha}^{\dagger}$ and $a_{\alpha}$ address the vibrational mode $\alpha$ with frequency $\Omega_{\alpha}$. The respective dimensionless displacement and momentum operators read 
		$Q_{\alpha} =  \frac{1}{\sqrt{2}}(a_{\alpha} + a^{\dagger}_{\alpha}), 
		P_{\alpha} =  \frac{-i}{\sqrt{2}} (a_{\alpha} - a^{\dagger}_{\alpha}).$ 
		
	In the next step, we expand the orbital energies, which enter the electronic part of the Hamiltonian, about the equilibrium geometry, $\textbf{R}_{0}$, of the reference state
		\begin{eqnarray}
		\epsilon_{m}(\textbf{R}) &=& \epsilon_{m}(\textbf{R}_{0}) + \sum_{\alpha} \left. \frac{\partial \epsilon_{m}(\textbf{R})}{\partial Q_{\alpha}} \right\vert_{\textbf{R}=\textbf{R}_{0}} \hspace*{-0.7cm} Q_{\alpha} + \dots \nonumber \\
					&\approx& \epsilon_{m} + \sum_{\alpha} \lambda_{m\alpha} Q_{\alpha}.
		\end{eqnarray}
	and similar for the Coulomb interaction 
		\begin{eqnarray}
		U_{mn}(\textbf{R}) &=& U_{mn}(\textbf{R}_{0}) + \sum_{\alpha} \left. \frac{\partial U_{mn}(\textbf{R})}{\partial Q_{\alpha}} \right\vert_{\textbf{R}=\textbf{R}_{0}} \hspace*{-0.7cm} Q_{\alpha} + \dots \nonumber \\
					&\approx& U_{mn} + \sum_{\alpha} W_{mn\alpha} Q_{\alpha}. \label{eq:Coulomb_expansion}
		\end{eqnarray}
	Notice that the form invariance of the Coulomb interaction is broken in (\ref{eq:Coulomb_expansion}), due to the fact that $U_{mn}(\textbf{R})$ represents an effective Coulomb interaction in a restricted model Hamiltonian.
	As a result, the Hamiltonian used to describe the molecular bridge is given by 
		\begin{eqnarray} \label{eq:H_final}
		H_{\text{S}} &=& \sum_{\alpha} \Omega_{\alpha} a_{\alpha}^{\dagger}a_{\alpha} +  \sum_{m} \epsilon_{m} (d_{m}^{\dagger}d_{m}-\delta_{m}) \nonumber \\ &&
				+  \sum_{mn} U_{mn} (d_{m}^{\dagger}d_{m}-\delta_{m})(d_{n}^{\dagger}d_{n}-\delta_{n}) \nonumber \\ &&
				+ \sum_{m\alpha} \lambda_{m\alpha} Q_{\alpha} (d_{m}^{\dagger}d_{m}-\delta_{m}) \nonumber \\ &&
				+ \sum_{mn\alpha} W_{mn\alpha} Q_{\alpha} (d_{m}^{\dagger}d_{m}-\delta_{m})(d_{n}^{\dagger}d_{n}-\delta_{n}). \nonumber  
		\end{eqnarray}
	It includes electron-electron interactions $U_{mn}$, electronic-vibrational coupling $\lambda_{m\alpha}$, and vibrationally dependent electron-electron interactions $W_{mn\alpha}$. 

\begin{figure}[tb]
	\includegraphics[width=0.5\textwidth]{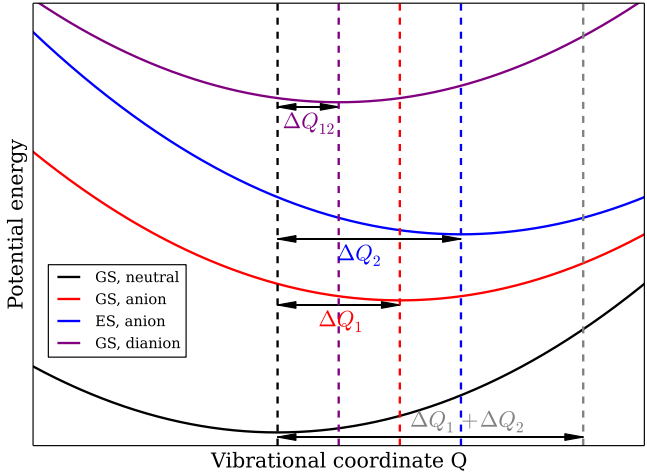}
	\caption{(Color online)\label{fig:PES}
			Scheme of potential energy surfaces of the molecular bridge corresponding to the ground state (GS) and first excited state (ES) of the neutral molecule, its anion and the dianion. The equilibrium geometry of the respective molecular states are marked by dashed vertical lines of the same color, the displacement of the equilibrium geometry with respect to the ground state of the neutral molecule is given by black arrows.}
	\end{figure}

	While the influence of electron-electron interactions and electronic-vibrational coupling on transport in molecular junctions have been studied in detail before \cite{Galperin07,cuevasscheer2010}, vibrationally dependent electron-electron interactions have so far only been considered in different contexts \cite{Schmalz2011, Pozner2014}. We first discuss their physical origin. The PESs of a generic one dimensional model for a molecular junction is depicted in Fig.\ \ref{fig:PES}, where the solid black line is associated with the ground state of the neutral molecule, the solid red line with the ground state of the respective anion, the solid blue line with the first excited state of the anion and the solid purple line with the ground state of the dianion. The minima of these PESs represent the equilibrium geometry of the nuclei, highlighted by dashed vertical lines. 
	The equilibrium geometry of the nuclei changes upon charging or excitation of the molecule. This constitutes one source of electronic-vibrational coupling. For example, the shift of the equilibrium geometry of the ground state of the anion with respect to the ground state of the neutral molecule is given by $\Delta Q_{1}=2 \lambda_{11}/\Omega_{1}$.
	Similarly, the equilibrium geometry of the first excited state of the anion is shifted by $\Delta Q_{2}=2 \lambda_{21}/\Omega_{1}$ with respect to the ground state of the neutral molecule. Without vibrationally dependent electron-electron interactions, the equilibrium position of the nuclei in the ground state of the dianion would be shifted by $\Delta Q_{12} = \Delta Q_{1}+\Delta Q_{2}=2 (\lambda_{11}+\lambda_{21})/\Omega_{1}$ with respect to the ground state of the neutral molecule, i.e.\ would be fixed by the parameters of the PES of the singly charged molecule. However, the equilibrium position of the nuclei of the dianion is not necessarily correctly described by this shift. The actual nuclear displacement can be characterized by an additional parameter via $\Delta Q_{12}=\Delta Q_{1}+\Delta Q_{2}+2 W_{121} /\Omega_{1}$. This shows that vibrationally dependent electron-electron interaction accounts for additional shifts of the equilibrium positions of the nuclei that occur if a state differs by the occupation of more than one single-particle state from the reference state (\emph{e.g.}\ $\ket{\Psi_{\text{ref}}(\mathbf{r;R})}$). 

	The molecule in a molecular junction is coupled to two electrodes, which we model by non-interacting electrons,
		\begin{eqnarray}
		H_{\text{L/R}} &=& \sum_{k\in\text{L/R}} \epsilon_{k} c_{k}^{\dagger}c_{k},
		\end{eqnarray}
	that are located on the left (L) and the right (R) electrode, respectively. The molecule-lead coupling can be described by 
		\begin{eqnarray}
		H_{\text{SL/SR}} &=& \sum_{m\in\text{M},k\in\text{L/R}} V_{mk} c_{k}^{\dagger}d_{m} + \text{H.c.},   
		\end{eqnarray}
	which allows for electron exchange processes between the molecule and the leads and determines the level-width function \begin{eqnarray}
	\Gamma_{\text{L/R} ij}(\epsilon) &=& 2\pi\sum_{k \in \text{L/R}}V_{\text{L/R} ik}V_{\text{L/R} jk}^* \delta(\epsilon-\epsilon_k). 
	\end{eqnarray}
	Throughout this article, the leads are modeled as semi-infinite tight binding chains with inter-site coupling strength $\beta$. The corresponding level-width function is given by
	\begin{eqnarray}
	 \Gamma_{\text{L/R} ij}(\epsilon) &=& \frac{\nu_{\text{L/R} i}\nu_{\text{L/R} j}^*}{\beta^2} \Theta(4\beta^2-x^2) \sqrt{4\beta^2-x^2},
	\end{eqnarray}
	with $x = \epsilon-\mu_{\text{L/R }}$ and the Heaviside step function $\Theta$ \cite{Cizek2004, Peskin2010, cuevasscheer2010}. Similar to $V_{ik}$ , the parameters $\nu_{\text{L/R} i}$ describe the coupling between the molecular state $i$ and the respective lead. 
	For simplicity, we neglect a dependency of the molecule-lead coupling on the nuclear degrees of freedom or the charge state of the molecule and assume, furthermore, a symmetric drop of bias voltage $\mu_\text{L}=-\mu_\text{R}=\Phi/2$.
	
	The overall Hamiltonian used to describe a single-molecule contact is given by 
		\begin{eqnarray}
		H &=& H_{\text{S}} + H_{\text{L}} + H_{\text{R}} + H_{\text{SL}} + H_{\text{SR}}.
		\end{eqnarray}

\subsection{Effective population-dependent electronic-vibrational coupling}\label{sec:Polaron}

	To understand the basic mechanism of vibrationally dependent electron-electron interactions, it is expedient to diagonalize the molecular part of the Hamiltonian $H_{\text{S}}$. To this end, we employ a generalized small polaron transformation \cite{Mahan81,Mitra04,Galperin06,Hartle} of the Hamiltonian 
		\begin{eqnarray}
		\overline{H} &=& \text{e}^{S} H \text{e}^{-S}  
		\end{eqnarray}
	with 
		\begin{eqnarray}
		S &=& - i \sum_{m\alpha} \frac{\lambda_{m\alpha}}{\Omega_{\alpha}}  ( d^{\dagger}_{m}d_{m} - \delta_{m} ) P_{\alpha} \\ &&
		- i \sum_{m<n\alpha} \frac{W_{mn\alpha}}{\Omega_{\alpha}}  ( d^{\dagger}_{m}d_{m} - \delta_{m} )  ( d^{\dagger}_{n}d_{n} - \delta_{n} ) P_{\alpha}.   \nonumber
		\end{eqnarray}
	and obtain 
		\begin{subequations} \label{eq:transformedHamiltonian}
		\begin{eqnarray}
		\overline{H} &=& \overline{H}_{\text{S}} + \overline{H}_{\text{B}} + \overline{H}_{\text{SB}} \\
		\overline{H}_{\text{S}} &=& \sum_{m} \overline{\epsilon}_{m} (d_{m}^{\dagger}d_{m} - \delta_{m} ) + \sum_{\alpha} \Omega_{\alpha} a^{\dagger}_{\alpha}a_{\alpha}\nonumber \\ &&
					+ \sum_{m<n} \overline{U}_{mn} ( d_{m}^{\dagger}d_{m} - \delta_{m} ) ( d^{\dagger}_{n}d_{n}  - \delta_{n} )  \\ 
		\overline{H}_{\text{SB}} &=& \sum_{m\in\text{M},k\in\text{L/R}} ( V_{mk} c_{k}^{\dagger}d_{m}X_{m} + \text{H.c.} )  , \\
		\overline{H}_{\text{B}} &=&  \overline{H}_{\text{L}} + \overline{H}_{\text{R}} = \sum_{k\in\text{L,R}} \epsilon_{k} c_{k}^{\dagger}c_{k} 
		\end{eqnarray}
		\end{subequations}
	The transformed Hamiltonian $\overline{H}$ has no explicit electronic-vibrational or vibrationally dependent electron-electron interaction terms. The effect of these interactions is subsumed in the polaron-shifted energy levels 
		\begin{eqnarray}
		\overline{\epsilon}_{m} &=& \epsilon_{m}-\sum_{\alpha}(\lambda_{m\alpha}^{2}/\Omega_{\alpha}), 
		\end{eqnarray}
	altered electron-electron interactions 
		\begin{eqnarray}  \label{eq:Ubar}
		\overline{U}_{mn} &=& U_{mn}-2\sum_{\alpha}(\lambda_{m\alpha} \lambda_{n\alpha}/\Omega_{\alpha})   \\ &&
		-  2 \sum_{\alpha} \frac{W_{mn\alpha}(\lambda_{m\alpha}+\lambda_{n\alpha})}{\Omega_{\alpha}} \nonumber
		- \sum_{\alpha} \frac{W_{mn\alpha}^{2}}{\Omega_{\alpha}}, 
		\end{eqnarray}
	and renormalized molecule-lead coupling matrix elements $ V_{mk}$ that are dressed by the shift operators 
		\begin{eqnarray}
		X_{m} &=& \text{exp}\left[i\sum_{\alpha} \left( \frac{\lambda_{m\alpha}}{\Omega_{\alpha}} +  \sum_{n \neq m} \frac{W_{mn\alpha}}{\Omega_{\alpha}} ( d^{\dagger}_{n}d_{n}  - \delta_{n} ) \right)
		P_{\alpha}\right]. \nonumber \\&&
		\end{eqnarray}
	In a similar way as the bare electronic-vibrational interaction $\lambda_{m\alpha}$ influences the electron-electron interaction ($U_{mn} \rightarrow U_{mn}-2\sum_{\alpha}(\lambda_{m\alpha} \lambda_{n\alpha}/\Omega_{\alpha})$), the vibrationally dependent electron-electron interaction leads to coupling terms proportional to $c_i^\dagger c_i c_j^\dagger c_j c_k^\dagger c_k$ ($i\neq j\neq k$) and $c_i^\dagger c_i c_j^\dagger c_j c_k^\dagger c_k c_l^\dagger c_l$ ($i\neq j\neq k\neq l$) in $\overline{H}$.
	To be consistent with the derivation of Eq.\ (\ref{eq:H_el_second}), we neglect those higher-order terms in Eqs.\ (\ref{eq:transformedHamiltonian}).
	
	Aside from an additional renormalization of the bare electron-electron interaction strengths, $U_{mn}\rightarrow\overline{U}_{mn}$, which induces no qualitatively new effects compared to the case with $W_{mn\alpha}=0$, vibrationally dependent electron-electron interactions affect, in particular, the structure of the shift operators $X_{m}$. Instead of $c$-numbered electronic-vibrational coupling strengths, $\lambda_{m\alpha}$, the shift operators $X_{m}$ involve effective electronic-vibrational coupling strengths 
		\begin{eqnarray} \label{eq:lambda_eff}
		\tilde{\lambda}_{m\alpha} &=& \lambda_{m\alpha} + \sum_{n\neq m} W_{mn\alpha} ( d^{\dagger}_{n}d_{n}  - \delta_{n} )
		\end{eqnarray}
	that include the electronic occupation operators $(d^{\dagger}_{n}d_{n}  - \delta_{n})$ with respect to all single-particle states but the $m$th one. As a result, the effective electronic-vibrational coupling strengths $\tilde{\lambda}_{m\alpha}$ depends on the population of the single-particle states and thus on the charge of the molecule. 
	Specifically, an electron that is transferred from the electrode to the $m$th state of the {\em neutral} molecule, couples with $\tilde{\lambda}_{m\alpha}=\lambda_{m\alpha}$ to the vibrational mode $\alpha$.
	For an electron that populates state $m$ of the {\em charged} molecule, where the occupation of the $n$th state differs from the neutral molecule, the effective electron-vibrational interaction is $\tilde{\lambda}_{m\alpha}= \lambda_{m\alpha} + W_{mn\alpha}$. 
	This implies the existence of different electronic-vibrational coupling strengths for the same electronic state, in the above example $\lambda_{m\alpha}$ and $\lambda_{m\alpha} + W_{mn\alpha}$. The coupling strength relevant for an electron entering/leaving the molecule depends on the exact population of all the other electronic states at the very moment the transport process takes place.
	An interpretation based on an averaged electronic-vibrational coupling strength $\lambda_{m\alpha} + W_{mn\alpha} \braket{ d^{\dagger}_{n}d_{n}  - \delta_{n} }$ does therefore not provide a correct description of the physics.
	A quantitative description requires an explicit calculation of the transport characteristics including vibrationally dependent electron-electron interactions, which will be the focus of Sec.\ \ref{sec:lambda_eff}.

	\subsection{Master equation approach}\label{sec:MEapproach}
	
	We simulate the transport properties of a single-molecule junction employing the well established Born-Markov master equation methodology \cite{May02,Mitra04,Lehmann04,Harbola2006,Volkovich2008,Hartle09,Hartle2010,Hartle2010b}. Thereby, the central object is the reduced density matrix $\rho$, which is obtained as the stationary solution of the equation of motion 
		\begin{eqnarray}
		\label{genfinalME}
		\partial_t \rho(t) &=& -i \left[ 
		\overline{H}_{\text{S}} , \rho(t) \right]  \\ &&
			- \int_{0}^{\infty} \text{d}\tau\, \text{tr}_{\text{B}}\lbrace \left[ \overline{H}_{\text{SB}} , \left[ \overline{H}_{\text{SB}}(\tau), \rho(t) \rho_{\text{B}} \right] \right] \rbrace , \nonumber
		\end{eqnarray}
	with 
		\begin{eqnarray}
		\overline{H}_{\text{SB}}(\tau) =  \text{e}^{-i(\overline{H}_{\text{S}}+\overline{H}_{\text{B}})\tau} \overline{H}_{\text{SB}} \text{e}^{i(\overline{H}_{\text{S}}+\overline{H}_{\text{B}})\tau} 
			\end{eqnarray}
	and $\rho_{\text{B}}$ being the equilibrium density matrix of the leads. Eq.\ (\ref{genfinalME}) can be obtained, e.g. using a second-order expansion of the exact Nakajima-Zwanzig equation \cite{Nakajima,Zwanzig} in the coupling $\overline{H}_{\text{SB}}$, including the so-called Markov approximation. 
	In the applications considered below, we will focus on the regime of resonant transport and weak molecule-lead coupling, where the Born-Markov master equation provides a correct description of the dominating transport processes. In this regime, where the molecule changes its charge state, the effects of vibrationally dependent electron-electron interaction are also expected to be most pronounced. 
	
	We evaluate the master equation (\ref{genfinalME}) for the steady state, focusing on model systems with two electronic states and a single vibrational mode, which is sufficient to show the generic effects of vibrationally dependent electron-electron interactions. In the calculations, we use basis functions $\vert n_{1}n_{2} \rangle\vert \nu \rangle$ ($n_{1},n_{2}\in\lbrace0,1\rbrace$) that represent the subspace of the electronic $\vert n_{1}n_{2} \rangle$ and the vibrational degrees of freedom $\vert \nu \rangle$ ($\nu\in\mathbb{N}_{0}$) in occupation number representations, respectively. The coefficients of the reduced density matrix are thus denoted by 
		\begin{eqnarray}
		\rho_{n_{1}n_{2},n'_{1}n'_{2}}^{\nu\nu'}	&\equiv& \langle n_{1}n_{2} \vert \rho^{\nu\nu'} \vert n'_{1}n'_{2} \rangle \nonumber \\
								&\equiv& \langle n_{1}n_{2} \vert \langle \nu \vert \rho \vert \nu' \rangle \vert n'_{1}n'_{2} \rangle.
		\end{eqnarray}
	The principal value terms in Eq.\ (\ref{genfinalME}) describe the renormalization of the molecular energy levels due to the coupling between the bridge and the leads \cite{Harbola2006,Hartle2010b,Hartle2014}. The importance of these terms has been investigated by {H\"artle} and Millis in a recent study of charge-transfer dynamics in a double quantum dot system \cite{Hartle2014}.
	For the systems considered in this work, where the single particle levels are well separated, these terms can be neglected. For the same reason we neglect vibrational coherences in the density matrix, which is a valid assumption for the systems considered here, which do not exhibit quasi-degeneracies and where the broadening due to molecule-lead coupling is small compared to the vibrational energies \cite{Hartle2010b, Schinabeck2014}.

	\subsection{Observables of interest}\label{sec:observables}
	
	For characterizing electron transport through a single-molecule junction, we analyze the electric current and the average vibrational excitation as a function of applied bias voltage $\Phi$. 

	Within the density matrix methodology outlined above, the expectation value of an observable $O$ can be calculated as
	\begin{eqnarray}
	 \braket{O} &=& \text{tr} \lbrace \rho O \rbrace = \sum_{n_1, n_2, \nu} \bra{n_1 n_2} \bra{\nu} \rho O \ket{\nu} \ket{n_1 n_2} 
	\end{eqnarray}
	Specifically, the excitation of the vibrational mode $\nu$ is given by \cite{Hartle2010b}
		\begin{eqnarray}
		&&\braket{a_\nu^\dagger a_\nu}_{H} = \nonumber \\
							&& \braket{a_\nu^\dagger a_\nu}_{\overline H}
							+2 \sum_{jk} \frac{\lambda_{j\nu}\lambda_{j\nu}}{\Omega_\nu^2}\braket{(d_j^\dagger d_j - \delta_j) (d_k^\dagger d_k - \delta_k)}_{\overline H}	\nonumber \\ &&
							+2 \sum_{i, j<k} \frac{\lambda_{i\nu}W_{jk\nu}}{\Omega_\nu^2} \braket{(d_i^\dagger d_i - \delta_i) (d_j^\dagger d_j - \delta_j) (d_k^\dagger d_k - \delta_k)}_{\overline H} \nonumber\\ &&
							+2 \sum_{i<j, k<l} \frac{W_{ij\nu}W_{kl\nu}}{\Omega_\nu^2} \langle (d_i^\dagger d_i - \delta_i) (d_j^\dagger d_j - \delta_j) (d_k^\dagger d_k - \delta_k) \nonumber \\ 
							&& \times (d_l^\dagger d_l - \delta_l)\rangle_{\overline H}
		\end{eqnarray}
	Thereby, the subscript $H$/$\overline H$ denotes the Hamiltonian, which is used to evaluate the respective expectation value.

	The number of electrons entering or leaving the lead $K$ ($K\in\lbrace \text{L,R}\rbrace$) per unit time determines the electronic current, 
		\begin{eqnarray}
		\label{firstcurrent}
		I_{K} &=& \langle \hat{I}_{K} \rangle_{H} = -2e \frac{\text{d}}{\text{d} t} \sum_{k\in K} \langle c^{\dagger}_{k} c_{k} \rangle_{\overline{H}} \\
			&=& 2ie \left[ \sum_{km} V_{mk} \langle c^{\dagger}_{k} d_{m} X_{m} \rangle_{\overline{H}} - \sum_{km} V_{mk}^{*} \langle d_{m}^{\dagger} X_{m}^{\dagger	} c_{k} \rangle_{\overline{H}} 	\right]. \nonumber
		\end{eqnarray}
	Here, the constant ($-e$) denotes the electron charge and the factor $2$ accounts for spin-degeneracy. 
	To second order in the molecule-lead coupling, the current through lead $K$ can be written as \cite{May02,Mitra04,Lehmann04,Harbola2006} 
		\begin{eqnarray}
		\label{gencurrentME}
		I_{K} &\approx& -i \int_{0}^{\infty}\text{d}\tau\, \text{tr}_{\text{S}+\text{B}}\lbrace \left[ \overline{H}_{\text{SB}}(\tau), \rho_{\text{B}}  \rho \right] \hat{I}_{K} \rbrace .
		\end{eqnarray}
	It is noted that this expression is current conserving, \emph{i.e.}\ $I_{\text{L}}=-I_{\text{R}}=I$.

\section{Results}\label{sec:Results}

		\begin{table*}[htb]
		 \caption{Parameters of the model systems investigated in this article. For all calculations, the temperature is set to $T=10$ K. All parameters are given in eV.}
 	  	\begin{center}
		\begin{tabular}{l||c|c|c|c|c|c|c|c|c|c|c|c}
		Model  & $\epsilon_1$ & $\epsilon_2$ & $ \nu_{\text{L }1}$ & $ \nu_{\text{R }1}$ & $ \nu_{\text{L }2}$ & $ \nu_{\text{R }2}$ & $\beta$ & $\hbar\Omega$ & $\lambda_1$ & $\lambda_2$ & $\overline{U}$ & $W$ \\ \hline\hline
		EFF    & $0.15$ & $0.3$ & $0.1$ & $0.1$ & $0.1$ & $0.1$ & $3$ & $0.1$ & $0.05$ & $0.05$ & $0.525$ & $0, \pm0.05$ \\
		STMSETUP  & $0.15$ & $0.2$ & $0.1$ & $0.01$ & $0.1$ & $0.01$ & $3$ & $0.1$ & $-0.05$ & $0.05$ & $1.05$ & $0, \pm0.05$ \\
		DARKST & $0.15$ & $0.2$ & $0.1$ & $0.1$ & $0.01$ & $0.01$ & $3$ & $0.1$ & $-0.05$ & $0.05$ & $0.05$ & $0, \pm0.05$ \\
		ASYMM  & $0.15$ & $0.3$ & $0.1$ & $0.03$ & $0.1$ & $0.03$ & $3$ & $0.1$ & $0.05$ & $0.05$ & $0.525$ & $0, \pm0.05$ 
		\end{tabular}
 		\end{center}
		\label{tab:parameters}
		\end{table*}

	To analyze the effect of vibrationally dependent electron-electron interactions, we consider a series of minimal models comprising two electronics levels and a single vibrational mode.
	The parameters of the models are summarized in Tab.\ \ref{tab:parameters}, where, for the sake of clarity, we dropped all vibrational indices and the electronic indices of $U$ and $W$. It is noted that all systems investigated exhibit relatively weak electronic-vibrational coupling strengths.
	
	The model parameters have been selected to study different aspects and mechanisms, including the manifestation of vibrationally dependent electron-electron interactions as an effective population-dependent electronic-vibrational coupling and its influence on vibrational excitation. Furthermore, signatures of  negative differential resistance and asymmetries in the gate voltage dependences of the current, even in symmetrically coupled molecular junctions, are investigated.

\subsection{Effective population-dependent electronic-vibrational coupling}\label{sec:lambda_eff}

	We start the analysis of the influence of vibrationally dependent electron-electron interaction by studying the transport properties of the generic model system EFF (see Tab.\ \ref{tab:parameters}) for different interaction strengths $W$. In this model, the electronic energy levels of the anion and the dianion are well separated, which allows for a better identification of the effect of vibrationally dependent electron-electron interaction. 
	As Eq.\ (\ref{eq:Ubar}) shows, vibrationally dependent electron-electron interaction renormalizes the Coulomb interaction strength, $U \rightarrow \overline{U}$, leading to a shift of the electronic resonances in the current-voltage characteristics.
	For better comparison, the value of $\overline{U}$ is fixed and the bare Coulomb interaction strength $U$ is adjusted such that the location of the electronic resonances coincides for all values of $W$.
	\begin{figure}[tb]
	\vspace*{0.75cm}
	\begin{tabular}{l}
	\hspace{-0.5cm}(a) \\[-0.75cm]
	\includegraphics[width=0.5\textwidth]{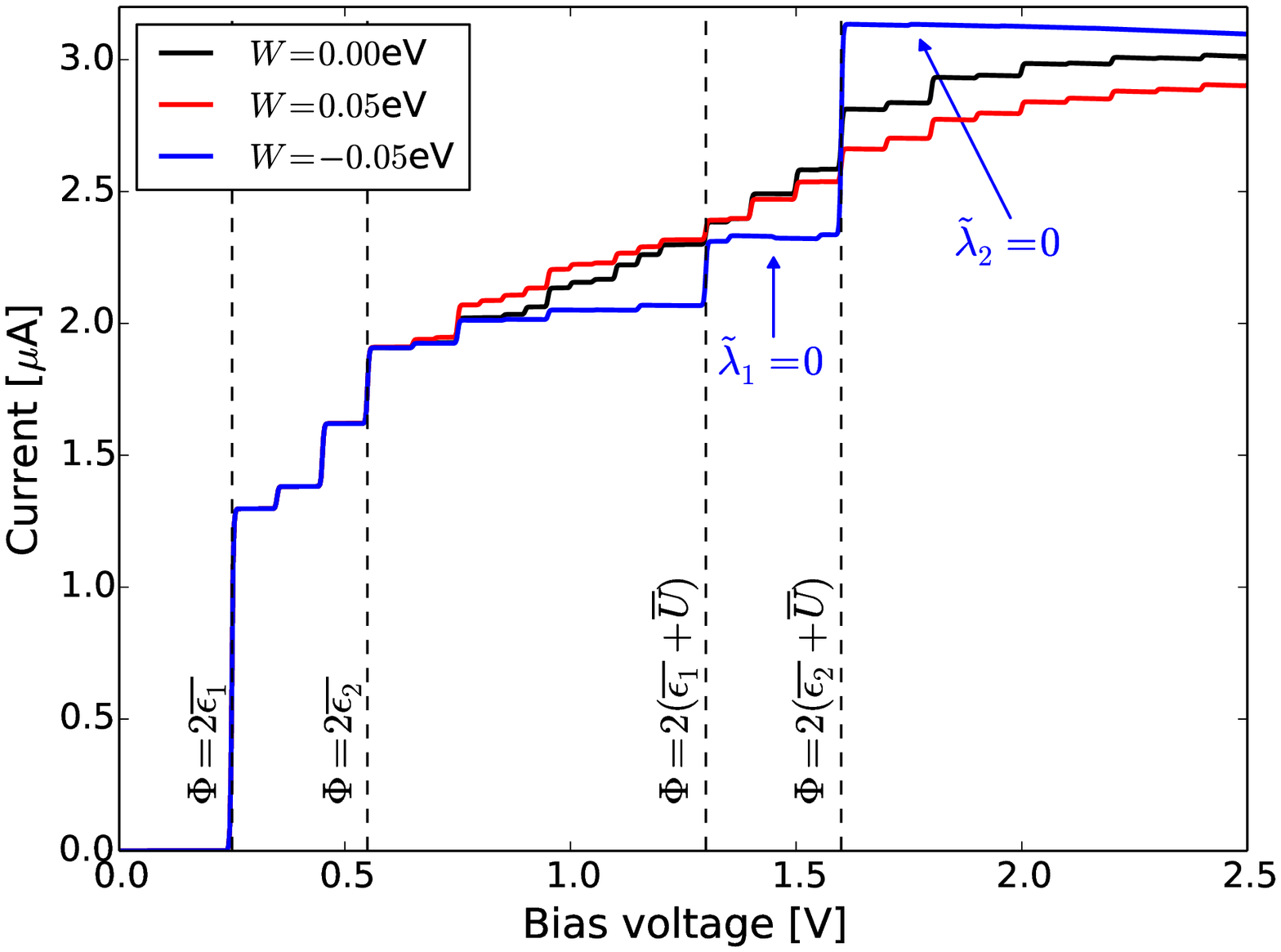}
	\\
	\hspace{-0.5cm}(b) \\[-0.65cm]
	\includegraphics[width=0.5\textwidth]{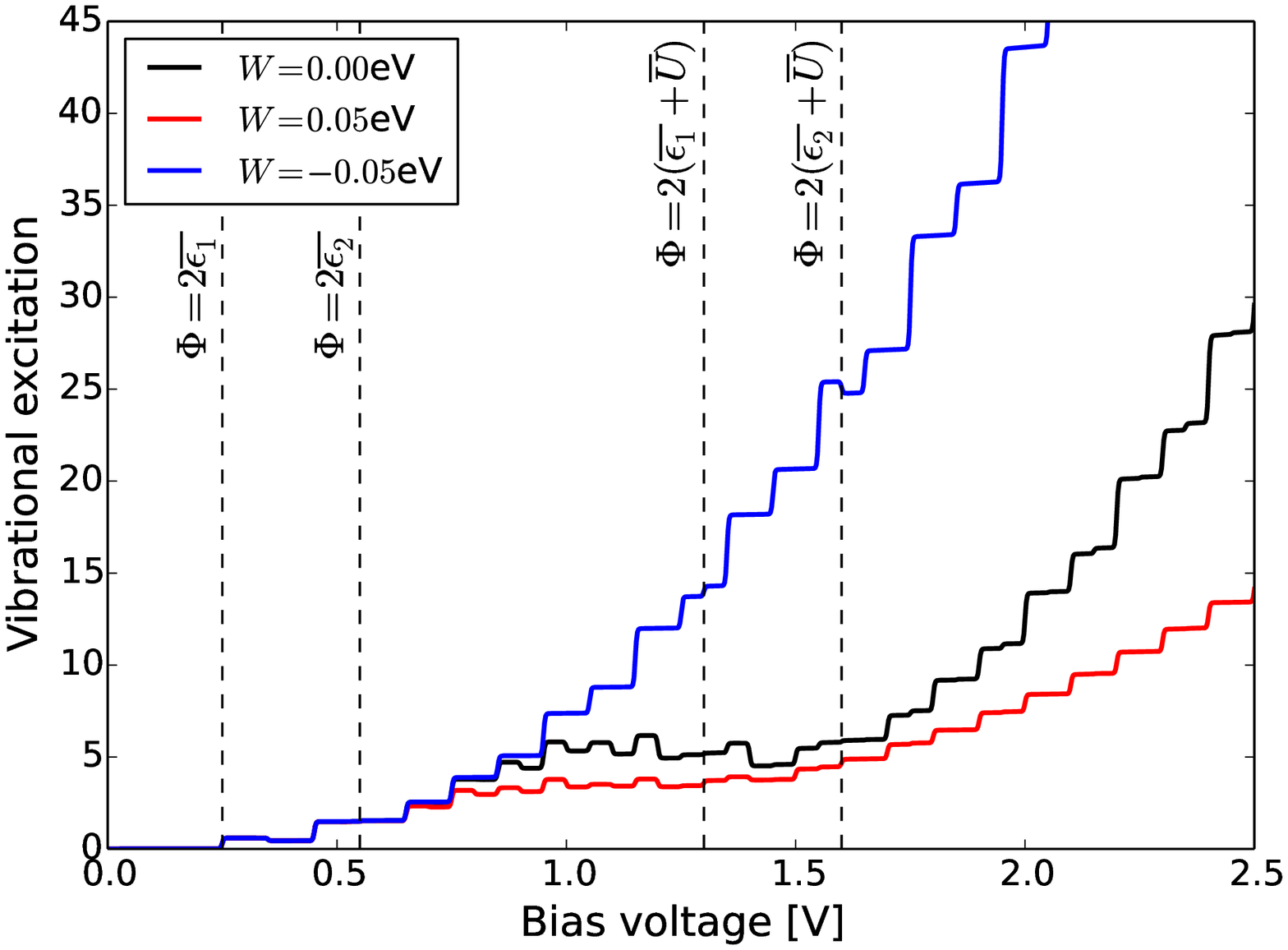}
	\end{tabular}
	\caption{(Color online) \label{fig:lambda_eff}
		(a) Current-voltage characteristics for the model system EFF. (b) Vibrational excitation as a function of bias voltage for the system EFF. The vertical dashed lines in both plots mark the onset of resonant transport through the corresponding molecular electronic states.}
	\end{figure}	

	Fig.\ \ref{fig:lambda_eff} shows the current and the vibrational excitation  as function of bias voltage $\Phi$ for model system EFF for $W=\pm0.05$ eV.  For comparison, also data obtained without vibrationally dependent electron-electron interaction, {\em i.e.}\ $W=0$ eV, are depicted. The results show that the vibrationally dependent electron-electron interaction has a significant effect for voltages $\Phi > 2\overline{\epsilon}_2$. 
	Depending on the sign of the interaction $W$ and the specific voltage, it results in a decrease or increase of the current compared to the system without interaction. Moreover, vibrational excitation is enhanced for $W=-0.05$ eV and reduced for $W=0.05$ eV.
 	
	These findings can be explained employing the effective electronic-vibrational coupling strength $\tilde{\lambda}_i$ introduced in Eq.\ (\ref{eq:lambda_eff}).
	We start by considering the case $W=+0.05$ eV. For low bias voltages, $\Phi \ll 2(\overline{\epsilon}_1+\overline{U})$, the current and the vibrational excitation agree with the noninteracting model. At these voltages, the features in the current-voltage and vibrational excitation characteristics are related to the opening of transport channels at energies $\overline{\epsilon}_{1/2}+m\Omega$ with $m\in \mathbb{Z}$. Populating both electronic states, $\overline{\epsilon}_1$ and $\overline{\epsilon}_2$, is not possible at these low voltages because the electrons coming from the leads do not have enough energy to overcome the Coulomb repulsion. 
	Accordingly, an electron impinging on the molecular bridge from the left electrode encounters a neutral molecule. Therefore, in this elementary charge transport step, the electron couples to the vibrational degrees of freedom with the coupling strengths $\tilde{\lambda}_i= \lambda_i +W \cdot 0 = \lambda_i$, which is independent of the vibrationally dependent electron-electron interaction $W$. As a consequence, the transport properties a low voltages are virtually identical to that of the system without  the vibrationally dependent electron-electron interaction.
	
	For higher voltages, $\Phi \gtrsim 2(\overline{\epsilon}_1+\overline{U})$, new features in the current and the vibrational excitation characteristic appear which are associated with the opening of transport channels at energies $\overline{\epsilon}_{1/2}+\overline{U}+m\Omega$ with $m\in \mathbb{Z}$ and correspond to transport channels where the electron impinging from the left electrode onto the molecule encounters a  
	singly occupied molecular bridge. As a consequence, in the elementary charge transport steps, the electron couples to the vibrational degrees of freedom $\tilde{\lambda}_i= \lambda_i +W \cdot 1=\lambda_i +W$. For $W=+0.05$ eV, the effective electronic-vibrational coupling is thus increased, which leads to two effects observed in Fig.\ \ref{fig:lambda_eff}: First, resonant transport processes associated with the absorption of vibrational energy are enhanced, resulting in an increased current for voltages below the onset of resonant transport involving the dianionic molecule, $\Phi < 2(\overline{\epsilon_2}+\overline{U})$.\cite{Hartle09, Hartle2010b} These processes also lead to a decreased vibrational excitation seen in the corresponding voltage regime in Fig.\ \ref{fig:lambda_eff}(b). Second, in the weak electronic-vibrational coupling regime considered here, an increased electronic-vibrational interaction gives rise to a decreased current and a diminished vibrational excitation beyond the onset of resonant transport involving the respective electronic state \cite{Ryndyk06, Semmelhack, Hartle09, Leturcq2009, Hartle2010b}, in this case the dianionic resonance, at voltages $\Phi > 2(\overline{\epsilon_2}+\overline{U})$.
	
	It is important to note that the current at these higher voltages includes both transport channels which couple with $\tilde{\lambda}_i = \lambda_i$ to the vibrations and which became active already at low bias voltages, and transport channels that couple with $\tilde{\lambda}_i = \lambda_i+W$. Although the average populations of the two electronic levels saturate at about $\braket{d_1^\dagger d_1} \approx \braket{d_2^\dagger d_2} \approx 0.5$ at high bias voltages, we want to stress that the above discussed transport behavior is not described correctly by considering transport with electronic-vibrational coupling strengths of $\tilde{\lambda}_i = \lambda_i+W\cdot 0.5$ corresponding to the average population, but needs to consider the populations of the elementary charge transport steps, which are zero or one, as discussed above.

	Next we consider the system with $W=-0.05$ eV. For low voltages, as discussed above, the transport is determined by the effective coupling $\tilde{\lambda}_i= \lambda_i +W \cdot 0$ and thus independent on $W$.
	At higher voltages, $\Phi \gtrsim 2(\overline{\epsilon}_1+\overline{U})$, however, the features in the current and the vibrational excitation are associated predominantly with transport processes, where an electron coming from the left electrode encounters a singly occupied molecular bridge resulting in an effective coupling of $\tilde{\lambda}_i= \lambda_i +W \cdot 1 = 0$ eV for the chosen parameters of the model, $\lambda=-W=0.05$ eV.
	This corresponds to a system where the electronic-vibrational coupling and the vibrationally dependent electron-electron interaction cancel each other. 
	Compared to the system with $W=0$ eV, this results in a smaller current for $2\overline{\epsilon_1} < e\Phi < 2(\overline{\epsilon_1}+\overline{U})$ and a larger current for $e\Phi > 2(\overline{\epsilon_2}+\overline{U})$. 
	At higher bias voltages the Franck-Condon blockade of the current is lifted, while the smaller current at lower voltages is due to the absence of transport processes associated with the absorption of vibrational energy.
	This is also reflected in the
	vibrational excitation characteristics (Fig.\ \ref{fig:lambda_eff}(b)), which for $W=-0.05$ eV exhibits a steady increase with voltage indicating the absence of processes that absorb vibrational energy. This increase of vibrational excitation is further enhanced because a decreased electronic-vibrational coupling leads to an increased vibrational excitation in the regime of weak electronic-vibrational coupling due to missing electron-hole pair creation processes that effectively cool the molecular bridge \cite{Hartle2011}. As a result, the largest average vibrational excitation is observed for the system $W=-0.05$ eV.

\subsection{Effect of vibrationally dependent electron-electron interactions in molecular junctions with left-right asymmetry }\label{sec:Vib_ex}

	Additional effect of vibrationally dependent electron-electron interactions arise in models with asymmetric coupling to left and right leads as is common, e.g., in STM setups. As an example, we consider the model system STMSETUP with parameters listed in Tab.\ \ref{tab:parameters}. Due to the difference in the coupling to the leads, the electronic states of the molecular bridge are completely occupied for positive bias voltages above the onset of resonant transport through the respective levels. For negative bias voltages, they are unoccupied. The anionic molecule can be in its electronic ground ($\overline{\epsilon}_1$ occupied, $\overline{\epsilon}_2$ unoccupied) or first excited state ($\overline{\epsilon}_1$ unoccupied, $\overline{\epsilon}_2$ occupied). Depending on the electronic configuration of the anion, the dianion is obtained by populating the first or the second electronic state. For the specific choice of the parameters, $\lambda_1=-\lambda_2=\pm W$, one of these channels for generating the dianion decouples from the vibrations ($\tilde\lambda_1=0$ eV or $\tilde\lambda_2=0$ eV, see below). This allows for a better identification of the effect of vibrationally dependent electron-electron interaction in the current and vibrational excitation characteristics.
	 
	\begin{figure}[tb]
	\vspace*{0.75cm}
	\begin{tabular}{l}
	\hspace{-0.5cm}(a) \\[-0.75cm]
	\includegraphics[width=0.5\textwidth]{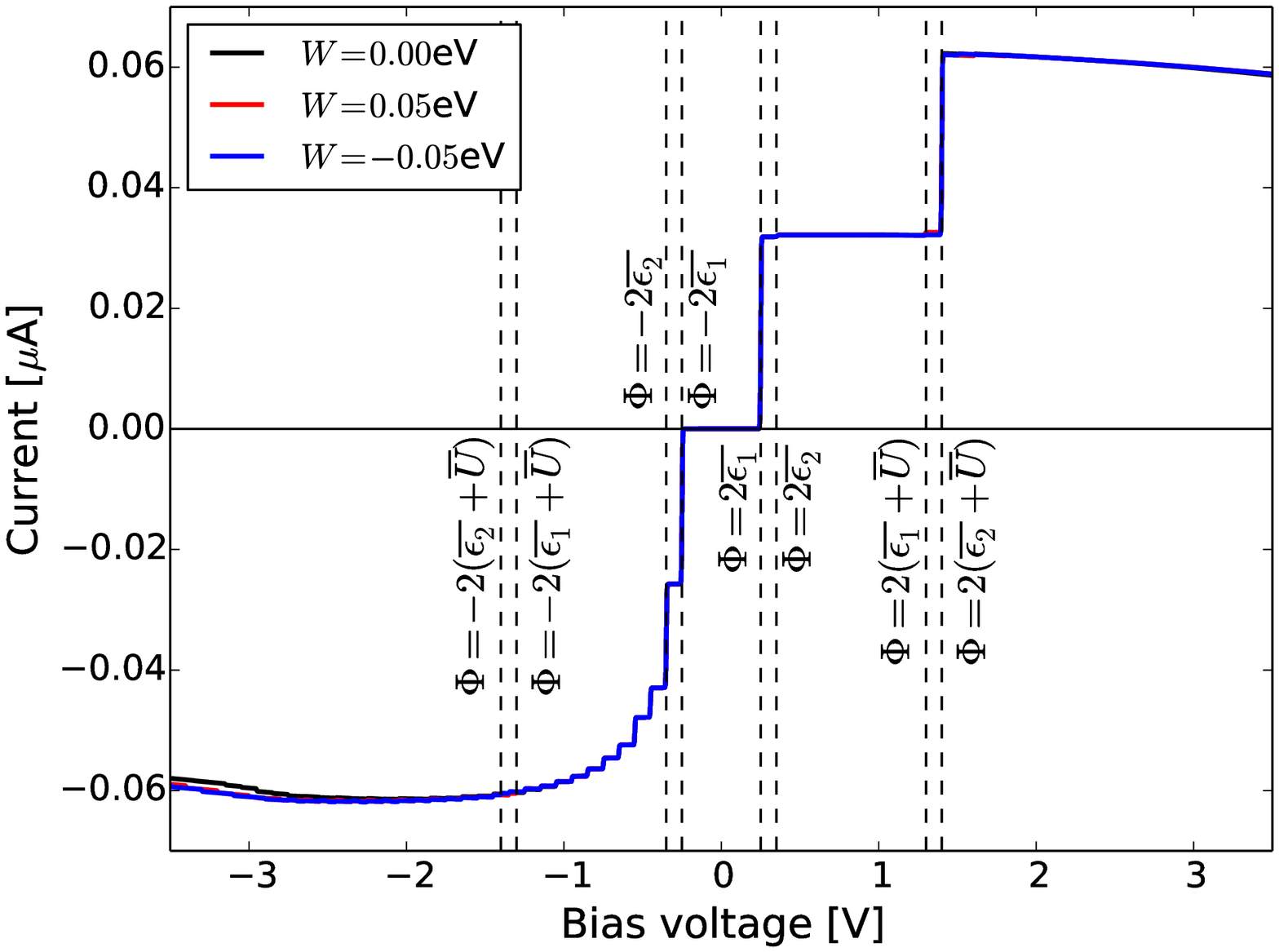}
	\\
	\hspace{-0.5cm}(b) \\[-0.65cm]
	\includegraphics[width=0.5\textwidth]{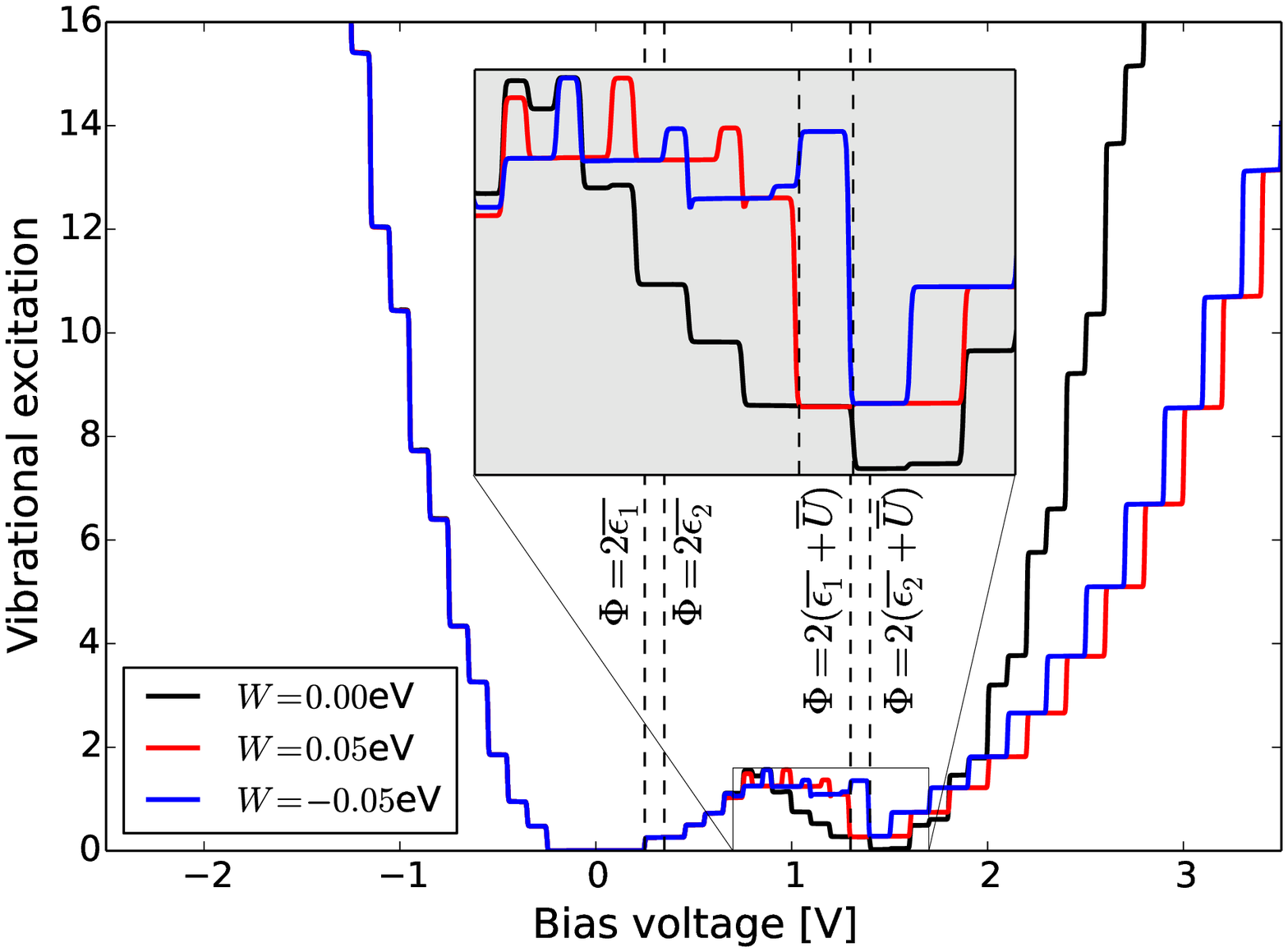}
	\end{tabular}
	\caption{(Color online) \label{fig:vib_excitation}
		(a) Current-voltage characteristics for the model system STMSETUP. (b) Vibrational excitation as a function of bias voltage for the system STMSETUP. The vertical dashed lines in both plots mark the onset of resonant transport through the corresponding molecular electronic states. The inset in panel (b) shows an enlargement of the region $\Phi \approx 2(\overline{\epsilon}_{1/2}+\overline{U})$.}
	\end{figure}	

	The current-voltage characteristics of model STMSETUP, depicted in Fig.\  \ref{fig:vib_excitation} (a), exhibits a pronounced  asymmetry with respect to bias polarity, which is a well known effect associated with the asymmetric coupling to the leads and the resulting dependence of the electronic population on bias polarity, \cite{Hartle2010b} but barely vary with $W$.  However, as shown in Fig.\ \ref{fig:vib_excitation}(b), there is a significant influence of the vibrationally dependent electron-electron interaction on the vibrational excitation for positive bias voltage. Two features are noteworthy:
	First, for $W\neq 0$ eV, there is a sudden decrease in vibrational excitation at voltages $\Phi = 2(\overline{\epsilon}_{1/2}+\overline{U})$, where the dianion becomes energetically accessible (see inset of Fig.\ \ref{fig:vib_excitation}(b)). Second, for high bias voltages, the vibrational excitation is significantly smaller compared to the $W=0$ case, independent of the sign of $W$. 

	These findings can be explained considering the population of the electronic states and the effective electronic-vibrational couplings  $\tilde\lambda_i$. Due to the asymmetric coupling to the leads, the molecule is mostly in the dianionic state, once the applied positive bias allows for double charging of the molecule. As a consequence, electrons populating the molecule from the left encounter an anionic molecule most of the times and transport processes with coupling $\tilde\lambda_{1/2} = \tilde\lambda_{1/2} + W\cdot 1$ to the vibrational degrees of freedom are dominant. Because one the effective couplings vanishes ($\tilde\lambda_1=0$ eV or $\tilde\lambda_2=0$ eV), the average vibrational excitation strongly decreases at the bias voltage where transport with $\tilde\lambda_{1/2}=0$ eV becomes energetically possible. 

	For large positive bias voltages, beyond the onset of transport involving the dianion, both molecular electronic states are almost always occupied. Accordingly, the current flowing across the molecule in this bias regime is determined by the transport processes that rely on the generation the dianion. For the system without vibrationally dependent electron-electron interaction, the transport through both electronic states couples with $\lambda$ to the nuclear degrees of freedom. Thus both transport channels cause a heating of the vibrational mode. For the systems with $W\neq0$ eV, on the other hand, only the transport channel through one of the electronic states couples to the vibrations with twice the coupling strength, $2\lambda$,  whereas the other decouples from the vibrations. As a stronger electronic-vibrational coupling leads to a decreased vibrational excitation in the regime of overall weak electronic-vibrational coupling (\emph{i.e.}\ $\lambda<\Omega$) \cite{Hartle09, Hartle2010b}, the systems with $W=\pm0.05$ eV exhibit on average a smaller number of vibrational quanta for large positive bias voltages.

\subsection{Negative differential resistance}\label{sec:NDR}

	Another effect introduced by vibrationally dependent electron-electron interaction is negative differential resistance (NDR), that is a decrease in current upon increase of bias voltage. To demonstrate this effect, we consider model system DARKST as specified in Tab.\ \ref{tab:parameters}. It consists of a strongly coupled electronic state $\overline\epsilon_1$, which is mainly responsible for the current flowing through the molecule, and a weaker coupled, or dark state $\overline\epsilon_2$, which mainly influences the transport properties of the main channel via vibrationally dependent electron-electron interaction. Again, the electron-electron interaction has been adapted such that the location of the electronic resonances coincides for the systems with $W=0, \pm 0.05$ eV. 
	\begin{figure}[tb]
	\begin{tabular}{l}
	\hspace{-0.5cm} \\[-1cm]
	\includegraphics[width=0.5\textwidth]{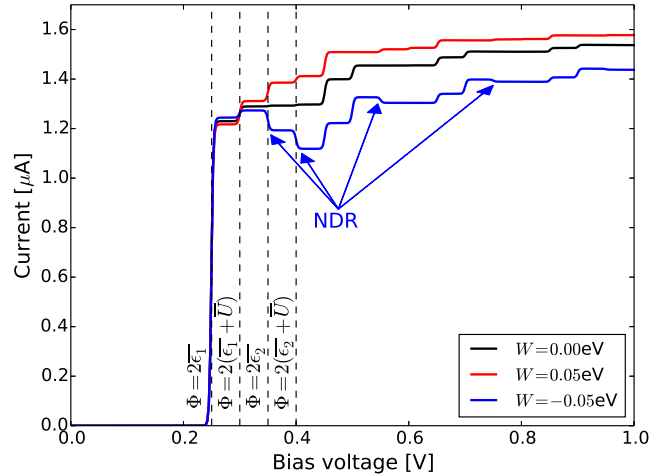}
	\end{tabular}
	\caption{(Color online) \label{fig:NDR}
		Current-voltage characteristics of model system DARKST. The vertical dashed lines mark the onset of resonant transport through molecular electronic states. }
	\end{figure}

	The current-voltage characteristics for the model system, depicted in Fig.\ \ref{fig:NDR}, exhibit one large step at $\Phi = 2\overline{\epsilon}_1$, marking the onset of resonant transport through $\overline{\epsilon}_1$, and additional, smaller vibrational features, which depend on the magnitude and sign of $W$. The influence of the second transport channel $\overline{\epsilon}_2$ is barely visible due to its weak coupling to the leads. For the case $W=0$, the main steps of the current are at $\Phi = 2(\overline{\epsilon}_1+n\Omega)$ and $\Phi = 2(\overline{\epsilon}_1+\overline{U}+n\Omega)$ with $n\in\mathbb{N}_0$. For $W=\pm0.05$ eV, however, there are also distinct features at $\Phi = 2(\overline{\epsilon}_2+n\Omega)$ and $\Phi = 2(\overline{\epsilon}_2+\overline{U}+n\Omega)$.
	While for $W=0.05$ eV the current is overall larger than for the reference system $W=0$, for $W=-0.05$ eV it is reduced. Furthermore, for $W=-0.05$ eV, the vibrational features  give rise to distinct decreases in the current upon increasing bias voltage. This NDR effect is marked by blue arrows in Fig.\ \ref{fig:NDR}.

	We first consider the results for $W=0.05$ eV. In this case, the effective couplings for transport through level $\overline{\epsilon}_1$ are $\tilde\lambda_1=-0.05$ eV and $\tilde\lambda_1=-0.05$ eV $+W = 0$ eV, respectively, depending on the occupation of the levels. For transport through level $\overline{\epsilon}_2$ the effective electronic-vibrational couplings are $\tilde\lambda_2=0.05$ eV and $\tilde\lambda_2=0.05$ eV $+W=0.1$ eV. Notice that the latter corresponds to an enhanced coupling for transport involving the dianion.
	With the population of $\overline{\epsilon}_2$ at biases $\Phi > 2\overline{\epsilon}_2$, the current flowing through $\overline{\epsilon}_1$ comprises transport channels that involve the dianion. As this transport path decouples from the vibrations, $\tilde\lambda_1=0$ eV, the current is increased compared to transport through $\overline{\epsilon}_1$ of the anionic molecule with $\tilde\lambda_1 = -0.05$ eV. 

	Next we study the current-voltage characteristics for $W=-0.05$ eV. In this case, the effective electronic-vibrational couplings for level $\overline{\epsilon}_1$ are $\tilde\lambda_1=-0.05$ eV and $\tilde\lambda_1=-0.05$ eV $+W = -0.1$ eV. Notice that transport through $\overline{\epsilon}_1$ involving the dianion corresponds to an enhanced electronic-vibrational coupling. For transport through level $\overline{\epsilon}_2$, the effective electronic-vibrational couplings are $\tilde\lambda_2=0.05$ eV and $\tilde\lambda_2=0.05$ eV $+W=0$ eV. 
	For $\Phi = 2(\overline{\epsilon}_1+n\Omega)$ and $\Phi = 2(\overline{\epsilon}_1+\overline{U}+n\Omega)$ we observe an increase in the current that is associated with the opening of new transport channels directly populating $\overline{\epsilon}_1$.
	For $\Phi = 2(\overline{\epsilon}_2+n\Omega)$ and $\Phi = 2(\overline{\epsilon}_2+\overline{U}+n\Omega)$, on the other hand, new transport channels directly populating level $\overline{\epsilon}_2$ become available. Consequently, transport processes involving the dianionic molecule become more important, thus increasing the significance of transport through $\overline{\epsilon}_1$ with enhanced electronic-vibrational coupling strength $\tilde\lambda_1= -0.1$ eV. As the transport through $\overline{\epsilon}_1$ dominates the current and an enhanced electronic-vibrational coupling results in a smaller current, a stepwise decrease of the current is observed at bias voltages $\Phi = 2(\overline{\epsilon}_2+n\Omega)$ and $\Phi = 2(\overline{\epsilon}_2+\overline{U}+n\Omega)$.
	Notice that this NDR effect results from the influence of the level $\overline{\epsilon}_2$ on the transport through $\overline{\epsilon}_1$ via the population dependent electronic-vibrational coupling $\tilde\lambda_1$.
	The NDR effect is therefore qualitatively different from similar NDR effects such as blocking state scenarios \cite{Hartle2010b}.

	Finally we want to remark that there are model systems, where the renormalization of the electron-electron interaction strength $U \rightarrow \overline{U}$ caused by the vibrationally dependent electron-electron interaction can also result in NDR. However, this effect does not appear in the model DARKST considered above.

\subsection{Asymmetries with respect to bias and gate voltage}\label{sec:Asymm}

	The vibrationally dependent electron-electron interaction leads to vibrational effects that are influenced by the electronic population of the molecular levels. As a consequence, vibrational features can change with bias polarity and gate voltage.
	In this section, we study these dependences based on conductance maps, that is the conductance as a function of bias and gate voltage. 
	Thereby we assume that the only effect of a gate voltage $\Phi_{\text{gate}}$ on the system is to shift the electronic energies of the noninteracting molecule, $\epsilon_i \rightarrow \epsilon_i + \Phi_{\text{gate}}$ \cite{Haertle2013}. 
	Other investigations of molecular junction transport based on conductance maps, also referred to as Coulomb diamonds or stability diagrams can be found, e.g., in \cite{Park2000, Kouwenhoven2001, Sapmaz05, Sapmaz2006, Chae2006, Thijssen2008, Leturcq2009, Huettel2009, Ballmann2012_2, Haertle2013}.

\begin{figure*}[tb]
		\centering
		\hspace{-0.5cm}
		\includegraphics[width=\textwidth]{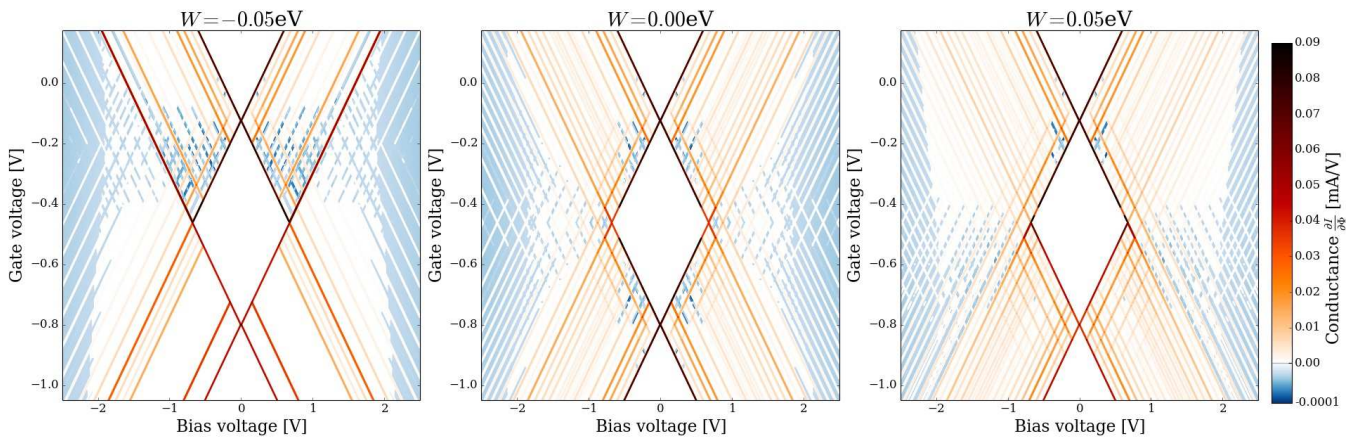}
		\caption{(Color online) \label{fig:CD_1}
			Conductance as a function of bias and gate voltage for the system EFF for $W=0, \pm0.05$ eV. Notice the different scale used for negative conductances.}
	\end{figure*}

	We start with model system EFF from Sec.\ \ref{sec:lambda_eff}, which is characterized by a symmetric coupling to the leads and vibrations. Fig.\ \ref{fig:CD_1} shows the conductance for coupling strengths of $W=0, \pm0.05$ eV.
	The conductance maps are dominated by the darker red lines, forming a diamond-shaped square in the center. These lines correspond to the onset of resonant transport through the electronic levels. Since the systems are corrected for the energy shift introduced by the vibrationally dependent electron-electron interaction, these features appear at the same positions for any value of $W$.
	Additionally, the conductance maps exhibit a distinct structure of less pronounced lines, which are associated with the onset of vibrationally coupled transport and differ in magnitude with $W$. 
	For $W=-0.05$ eV, fewer vibrational lines are visible and for $W=+0.05$ eV, the vibrational structures are observed over a wider range of voltages compared to the $W=0$ eV case.
	For $W= 0$ eV, the conductance maps exhibits three symmetries with respect to the transformations $\Phi_{\text{bias}} \rightarrow -\Phi_{\text{bias}}$,  $\Phi_{\text{gate}} \rightarrow -\Phi_{\text{gate}}-(\overline{\epsilon}_1+\overline{\epsilon}_2+\overline{U})$ and consequently also with respect to $\Phi_{\text{bias}}, \Phi_{\text{gate}}  \rightarrow -\Phi_{\text{bias}},  -\Phi_{\text{gate}} - (\overline{\epsilon}_1+\overline{\epsilon}_2+\overline{U})$. 
	The first is due to the symmetric coupling of the molecule to the leads and is unaffected by the vibrationally dependent electron-electron interaction.
	The second and third symmetry are broken for $W\neq 0$ eV.
	Lastly, the data show blue patterns, corresponding to small negative differential resistance, located at high bias voltages but also in areas close to the onset of resonant transport. 

	These findings can be rationalized as follows. The location of the lines related to electronic and vibronic transport are unchanged by $W$ because the vibrationally dependent electron-electron interaction does not alter the vibrational energy. 
	For $W=-0.05$ eV, fewer vibrational transport channels exist as transport involving the dianionic resonance effectively decouples from the nuclear degrees of freedom. For $W=+0.05$ eV, transport including the dianionic molecule couples with twice the strength to the nuclear displacement such that processes including several vibrational quanta are more pronounced than in the $W=0$ eV case. The symmetry with respect to the gate voltage is lifted because
	the vibrational effects depend on the electronic population of the molecule. This leads to a change of the effective electronic-vibrational coupling upon variation in gate voltage, and hence to a different vibrational structure. 
	The origin of the NDR for high bias voltages is the finite bandwidth of the leads, which are modeled as semi-infinite chains. 
	The NDR for the model system with $W=0.00$ eV for gate voltages around $-0.2$ V, $-0.475$ V and $-0.7$ V is caused by the influence of the vibrational nonequilibrium state on the transport properties of the junction as discussed in Ref.\ \onlinecite{Haertle11}. The change in the NDR structure for $W\neq 0$ eV is related to the change in the effective electronic-vibrational coupling for transport involving the dianion, resulting in an altered vibrational nonequilibrium state and the effect discussed in Sec.\ \ref{sec:NDR}.

\begin{figure*}[htb]
		\centering
		\hspace{-0.5cm}
		\includegraphics[width=\textwidth]{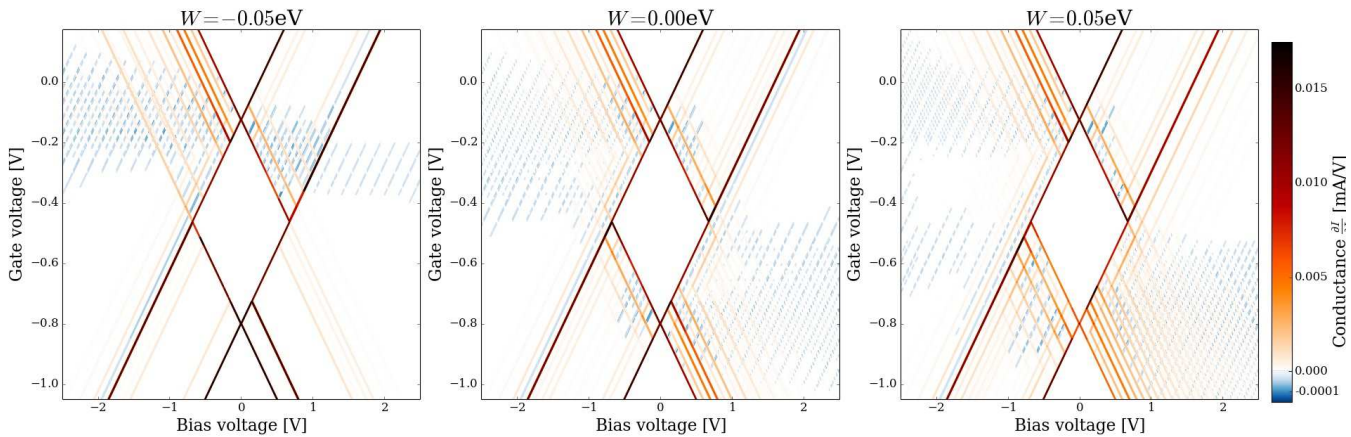}
		\caption{(Color online) \label{fig:CD_2}
			Conductance as a function of bias and gate voltage for the system ASYMM for $W=0, \pm0.05$ eV. Notice the different scaling for negative conductances.}
	\end{figure*}

	Next, we consider the model system ASYMM with asymmetric molecule-lead coupling, which allows to study the relation between the vibrationally dependent electron-electron interaction and bias polarity. The corresponding conductance maps for $W=0,\pm0.05$ eV are shown in Fig.\ \ref{fig:CD_2}. 	
	As a consequence of its asymmetric coupling to the leads, the electronic population of this model is sensitive to the polarity of the applied bias and so is the influence of the vibrationally dependent electron-electron interaction. As in the previous model, we observe a change in the vibrational structure of the conductance map upon a variation in $W$. For $W=-0.05$ eV, the lines corresponding to vibrational transport are less pronounced, in particular for gate voltages below $\approx -0.45$ V only few vibrational features are present. For $W=+0.05$ eV, the vibrational features are more pronounced and are observed for all gate voltages.
	The data also exhibit weak NDR effects, which are influenced by the vibrationally dependent electron-electron interaction.
	Due to the asymmetric coupling to the leads, there is no symmetry in the conductance maps with respect to bias polarity, $\Phi_{\text{bias}} \rightarrow -\Phi_{\text{bias}}$.
	Remarkably, also the symmetry of the conductance map under the transformation $\Phi_{\text{bias}}, \Phi_{\text{gate}}  \rightarrow -\Phi_{\text{bias}},  -\Phi_{\text{gate}} - (\overline{\epsilon}_1+\overline{\epsilon}_2+\overline{U})$, is broken for $W\neq0$. In the $W=0$ eV case, this symmetry is a consequence of the fact that both bias and gate voltage change the average population of the electronic states in a similar way and that there is no distinction between transport involving the anion or the dianion. Including vibrationally dependent electron-electron interaction, transport involving the anion or the dianion is no longer equivalent. As a result the transport mechanism depends on the total charge of the molecule and, therefore, this symmetry of the conductance map is broken. It is interesting to note that most of the conductance maps measured in experiment display asymmetries  \cite{Park2000, Kouwenhoven2001, Sapmaz05, Sapmaz2006, Huettel2009}

\section{Conclusion}
	
	We have investigated the effect of vibrationally dependent electron-electron interactions in single-molecule junctions. 
	The additional interaction is a result of the dependence of the Coulomb interaction on the nuclear displacement and accounts for the fact that vibronic transport processes depend on the charge state of the molecule, i.e.\ are different for transport through, e.g., an anionic or dianionic state of the molecule. 
	Employing a generalized small polaron transform, we have shown that vibrationally dependent electron-electron interaction results in an effective electronic-vibrational coupling, which depends on the electronic occupation, and can be used to rationalize the effects on charge transport.
	
	Employing a master equation approach, we have analyzed the basic mechanism  and the manifestations of vibrationally dependent electron-electron interactions in single-molecule junctions. 
	Depending on the strength an the sign of the interaction it may result in a significant alteration of the transport characteristics and the vibrational nonequilibrium excitation and may cause NDR. For selected values of the interaction strength, the interplay between electronic-vibrational interaction and vibrationally dependent electron-electron interaction can also lead to regimes where electronic-vibrational coupling is effectively switched off. In junctions with asymmetric molecule-lead coupling, vibrationally dependent electron-electron interaction may cause a strong dependence of the vibrational nonequilibrium excitation on the bias polarity. 
	Finally, vibrationally dependent electron-electron interaction can give rise to asymmetries of conductance maps related to the different description of transport involving the anionic and the dianionic molecule. The latter
	finding may be of particular interest in the context of experimental results, which have the tendency to display asymmetries \cite{Park2000, Kouwenhoven2001, Sapmaz05, Sapmaz2006, Huettel2009}.

\section{Acknowledgments}
	We thank C.\ Schinabeck and P.\ B.\ Coto for helpful discussions.
	This work has been supported by the
	German-Israeli Foundation for Scientific Development (GIF) and the
	Deutsche Forschungsgemeinschaft (DFG) through the DFG-Cluster of
	Excellence 'Engineering of Advanced Materials', SFB 953,
	and a research grant. The generous allocation
	of computing time by the computing centers in Erlangen
	(RRZE) and Munich (LRZ) is gratefully acknowledged.
	RH was supported by the Alexander von Humdoldt foundation via a Feodor-Lynen research fellowship.

\bibliography{my}

\begin{thebibliography}{89}%
\makeatletter
\providecommand \@ifxundefined [1]{%
 \@ifx{#1\undefined}
}%
\providecommand \@ifnum [1]{%
 \ifnum #1\expandafter \@firstoftwo
 \else \expandafter \@secondoftwo
 \fi
}%
\providecommand \@ifx [1]{%
 \ifx #1\expandafter \@firstoftwo
 \else \expandafter \@secondoftwo
 \fi
}%
\providecommand \natexlab [1]{#1}%
\providecommand \enquote  [1]{``#1''}%
\providecommand \bibnamefont  [1]{#1}%
\providecommand \bibfnamefont [1]{#1}%
\providecommand \citenamefont [1]{#1}%
\providecommand \href@noop [0]{\@secondoftwo}%
\providecommand \href [0]{\begingroup \@sanitize@url \@href}%
\providecommand \@href[1]{\@@startlink{#1}\@@href}%
\providecommand \@@href[1]{\endgroup#1\@@endlink}%
\providecommand \@sanitize@url [0]{\catcode `\\12\catcode `\$12\catcode
  `\&12\catcode `\#12\catcode `\^12\catcode `\_12\catcode `\%12\relax}%
\providecommand \@@startlink[1]{}%
\providecommand \@@endlink[0]{}%
\providecommand \url  [0]{\begingroup\@sanitize@url \@url }%
\providecommand \@url [1]{\endgroup\@href {#1}{\urlprefix }}%
\providecommand \urlprefix  [0]{URL }%
\providecommand \Eprint [0]{\href }%
\providecommand \doibase [0]{http://dx.doi.org/}%
\providecommand \selectlanguage [0]{\@gobble}%
\providecommand \bibinfo  [0]{\@secondoftwo}%
\providecommand \bibfield  [0]{\@secondoftwo}%
\providecommand \translation [1]{[#1]}%
\providecommand \BibitemOpen [0]{}%
\providecommand \bibitemStop [0]{}%
\providecommand \bibitemNoStop [0]{.\EOS\space}%
\providecommand \EOS [0]{\spacefactor3000\relax}%
\providecommand \BibitemShut  [1]{\csname bibitem#1\endcsname}%
\let\auto@bib@innerbib\@empty
\bibitem [{\citenamefont {Galperin}\ \emph {et~al.}(2007)\citenamefont
  {Galperin}, \citenamefont {Ratner},\ and\ \citenamefont
  {Nitzan}}]{Galperin07}%
  \BibitemOpen
  \bibfield  {author} {\bibinfo {author} {\bibfnamefont {M.}~\bibnamefont
  {Galperin}}, \bibinfo {author} {\bibfnamefont {M.~A.}\ \bibnamefont
  {Ratner}}, \ and\ \bibinfo {author} {\bibfnamefont {A.}~\bibnamefont
  {Nitzan}},\ }\href@noop {} {\bibfield  {journal} {\bibinfo  {journal} {J.
  Phys.: Condens. Matter}\ }\textbf {\bibinfo {volume} {19}},\ \bibinfo {pages}
  {103201} (\bibinfo {year} {2007})}\BibitemShut {NoStop}%
\bibitem [{\citenamefont {Wu}\ \emph {et~al.}(2004)\citenamefont {Wu},
  \citenamefont {Nazin}, \citenamefont {Chen}, \citenamefont {Qiu},\ and\
  \citenamefont {Ho}}]{WuNazinHo04}%
  \BibitemOpen
  \bibfield  {author} {\bibinfo {author} {\bibfnamefont {S.~W.}\ \bibnamefont
  {Wu}}, \bibinfo {author} {\bibfnamefont {G.~V.}\ \bibnamefont {Nazin}},
  \bibinfo {author} {\bibfnamefont {X.}~\bibnamefont {Chen}}, \bibinfo {author}
  {\bibfnamefont {X.~H.}\ \bibnamefont {Qiu}}, \ and\ \bibinfo {author}
  {\bibfnamefont {W.}~\bibnamefont {Ho}},\ }\href@noop {} {\bibfield  {journal}
  {\bibinfo  {journal} {Phys. Rev. Lett.}\ }\textbf {\bibinfo {volume} {93}},\
  \bibinfo {pages} {236802} (\bibinfo {year} {2004})}\BibitemShut {NoStop}%
\bibitem [{\citenamefont {LeRoy}\ \emph {et~al.}(2004)\citenamefont {LeRoy},
  \citenamefont {Lemay}, \citenamefont {Kong},\ and\ \citenamefont
  {Dekker}}]{LeRoy}%
  \BibitemOpen
  \bibfield  {author} {\bibinfo {author} {\bibfnamefont {B.~J.}\ \bibnamefont
  {LeRoy}}, \bibinfo {author} {\bibfnamefont {S.~G.}\ \bibnamefont {Lemay}},
  \bibinfo {author} {\bibfnamefont {J.}~\bibnamefont {Kong}}, \ and\ \bibinfo
  {author} {\bibfnamefont {C.}~\bibnamefont {Dekker}},\ }\href@noop {}
  {\bibfield  {journal} {\bibinfo  {journal} {Nature}\ }\textbf {\bibinfo
  {volume} {432}},\ \bibinfo {pages} {371} (\bibinfo {year}
  {2004})}\BibitemShut {NoStop}%
\bibitem [{\citenamefont {Yu}\ \emph {et~al.}(2004)\citenamefont {Yu},
  \citenamefont {Keane}, \citenamefont {Ciszek}, \citenamefont {Cheng},
  \citenamefont {Stewart}, \citenamefont {Tour},\ and\ \citenamefont
  {Natelson}}]{Natelson04}%
  \BibitemOpen
  \bibfield  {author} {\bibinfo {author} {\bibfnamefont {L.~H.}\ \bibnamefont
  {Yu}}, \bibinfo {author} {\bibfnamefont {Z.~K.}\ \bibnamefont {Keane}},
  \bibinfo {author} {\bibfnamefont {J.~W.}\ \bibnamefont {Ciszek}}, \bibinfo
  {author} {\bibfnamefont {L.}~\bibnamefont {Cheng}}, \bibinfo {author}
  {\bibfnamefont {M.~P.}\ \bibnamefont {Stewart}}, \bibinfo {author}
  {\bibfnamefont {J.~M.}\ \bibnamefont {Tour}}, \ and\ \bibinfo {author}
  {\bibfnamefont {D.}~\bibnamefont {Natelson}},\ }\href@noop {} {\bibfield
  {journal} {\bibinfo  {journal} {Phys. Rev. Lett.}\ }\textbf {\bibinfo
  {volume} {93}},\ \bibinfo {pages} {266802} (\bibinfo {year}
  {2004})}\BibitemShut {NoStop}%
\bibitem [{\citenamefont {Pasupathy}\ \emph {et~al.}(2005)\citenamefont
  {Pasupathy}, \citenamefont {Park}, \citenamefont {Chang}, \citenamefont
  {Soldatov}, \citenamefont {Lebedkin}, \citenamefont {Bialczak}, \citenamefont
  {Grose}, \citenamefont {Donev}, \citenamefont {Sethna}, \citenamefont
  {Ralph},\ and\ \citenamefont {McEuen}}]{Pasupathy05}%
  \BibitemOpen
  \bibfield  {author} {\bibinfo {author} {\bibfnamefont {A.~N.}\ \bibnamefont
  {Pasupathy}}, \bibinfo {author} {\bibfnamefont {J.}~\bibnamefont {Park}},
  \bibinfo {author} {\bibfnamefont {C.}~\bibnamefont {Chang}}, \bibinfo
  {author} {\bibfnamefont {A.~V.}\ \bibnamefont {Soldatov}}, \bibinfo {author}
  {\bibfnamefont {S.}~\bibnamefont {Lebedkin}}, \bibinfo {author}
  {\bibfnamefont {R.~C.}\ \bibnamefont {Bialczak}}, \bibinfo {author}
  {\bibfnamefont {J.~E.}\ \bibnamefont {Grose}}, \bibinfo {author}
  {\bibfnamefont {L.~A.~K.}\ \bibnamefont {Donev}}, \bibinfo {author}
  {\bibfnamefont {J.~P.}\ \bibnamefont {Sethna}}, \bibinfo {author}
  {\bibfnamefont {D.~C.}\ \bibnamefont {Ralph}}, \ and\ \bibinfo {author}
  {\bibfnamefont {P.~L.}\ \bibnamefont {McEuen}},\ }\href@noop {} {\bibfield
  {journal} {\bibinfo  {journal} {Nano Lett.}\ }\textbf {\bibinfo {volume}
  {5}},\ \bibinfo {pages} {203} (\bibinfo {year} {2005})}\BibitemShut {NoStop}%
\bibitem [{\citenamefont {Sapmaz}\ \emph {et~al.}(2006)\citenamefont {Sapmaz},
  \citenamefont {Jarillo-Herrero}, \citenamefont {Blanter}, \citenamefont
  {Dekker},\ and\ \citenamefont {van~der Zant}}]{Sapmaz2006}%
  \BibitemOpen
  \bibfield  {author} {\bibinfo {author} {\bibfnamefont {S.}~\bibnamefont
  {Sapmaz}}, \bibinfo {author} {\bibfnamefont {P.}~\bibnamefont
  {Jarillo-Herrero}}, \bibinfo {author} {\bibfnamefont {Y.~M.}\ \bibnamefont
  {Blanter}}, \bibinfo {author} {\bibfnamefont {C.}~\bibnamefont {Dekker}}, \
  and\ \bibinfo {author} {\bibfnamefont {H.~S.~J.}\ \bibnamefont {van~der
  Zant}},\ }\href@noop {} {\bibfield  {journal} {\bibinfo  {journal} {Phys.
  Rev. Lett.}\ }\textbf {\bibinfo {volume} {96}},\ \bibinfo {pages} {026801}
  (\bibinfo {year} {2006})}\BibitemShut {NoStop}%
\bibitem [{\citenamefont {Thijssen}\ \emph {et~al.}(2006)\citenamefont
  {Thijssen}, \citenamefont {Djukic}, \citenamefont {Otte}, \citenamefont
  {Bremmer},\ and\ \citenamefont {van Ruitenbeek}}]{Thijssen06}%
  \BibitemOpen
  \bibfield  {author} {\bibinfo {author} {\bibfnamefont {W.~H.~A.}\
  \bibnamefont {Thijssen}}, \bibinfo {author} {\bibfnamefont {D.}~\bibnamefont
  {Djukic}}, \bibinfo {author} {\bibfnamefont {A.~F.}\ \bibnamefont {Otte}},
  \bibinfo {author} {\bibfnamefont {R.~H.}\ \bibnamefont {Bremmer}}, \ and\
  \bibinfo {author} {\bibfnamefont {J.~M.}\ \bibnamefont {van Ruitenbeek}},\
  }\href@noop {} {\bibfield  {journal} {\bibinfo  {journal} {Phys. Rev. Lett.}\
  }\textbf {\bibinfo {volume} {97}},\ \bibinfo {pages} {226806} (\bibinfo
  {year} {2006})}\BibitemShut {NoStop}%
\bibitem [{\citenamefont {Parks}\ \emph {et~al.}(2007)\citenamefont {Parks},
  \citenamefont {Champagne}, \citenamefont {Hutchison}, \citenamefont
  {Flores-Torres}, \citenamefont {Abruna},\ and\ \citenamefont
  {Ralph}}]{Parks07}%
  \BibitemOpen
  \bibfield  {author} {\bibinfo {author} {\bibfnamefont {J.~J.}\ \bibnamefont
  {Parks}}, \bibinfo {author} {\bibfnamefont {A.~R.}\ \bibnamefont
  {Champagne}}, \bibinfo {author} {\bibfnamefont {G.~R.}\ \bibnamefont
  {Hutchison}}, \bibinfo {author} {\bibfnamefont {S.}~\bibnamefont
  {Flores-Torres}}, \bibinfo {author} {\bibfnamefont {H.~D.}\ \bibnamefont
  {Abruna}}, \ and\ \bibinfo {author} {\bibfnamefont {D.~C.}\ \bibnamefont
  {Ralph}},\ }\href@noop {} {\bibfield  {journal} {\bibinfo  {journal} {Phys.
  Rev. Lett.}\ }\textbf {\bibinfo {volume} {99}},\ \bibinfo {pages} {026601}
  (\bibinfo {year} {2007})}\BibitemShut {NoStop}%
\bibitem [{\citenamefont {{B\"ohler}}\ \emph {et~al.}(2007)\citenamefont
  {{B\"ohler}}, \citenamefont {Edtbauer},\ and\ \citenamefont
  {Scheer}}]{Boehler07}%
  \BibitemOpen
  \bibfield  {author} {\bibinfo {author} {\bibfnamefont {T.}~\bibnamefont
  {{B\"ohler}}}, \bibinfo {author} {\bibfnamefont {A.}~\bibnamefont
  {Edtbauer}}, \ and\ \bibinfo {author} {\bibfnamefont {E.}~\bibnamefont
  {Scheer}},\ }\href@noop {} {\bibfield  {journal} {\bibinfo  {journal} {Phys.
  Rev. B}\ }\textbf {\bibinfo {volume} {76}},\ \bibinfo {pages} {125432}
  (\bibinfo {year} {2007})}\BibitemShut {NoStop}%
\bibitem [{\citenamefont {{de Leon}}\ \emph {et~al.}(2008)\citenamefont {{de
  Leon}}, \citenamefont {Liang}, \citenamefont {Gu},\ and\ \citenamefont
  {Park}}]{Leon2008}%
  \BibitemOpen
  \bibfield  {author} {\bibinfo {author} {\bibfnamefont {N.~P.}\ \bibnamefont
  {{de Leon}}}, \bibinfo {author} {\bibfnamefont {W.}~\bibnamefont {Liang}},
  \bibinfo {author} {\bibfnamefont {Q.}~\bibnamefont {Gu}}, \ and\ \bibinfo
  {author} {\bibfnamefont {H.}~\bibnamefont {Park}},\ }\href@noop {} {\bibfield
   {journal} {\bibinfo  {journal} {Nano Lett.}\ }\textbf {\bibinfo {volume}
  {8}},\ \bibinfo {pages} {2963} (\bibinfo {year} {2008})}\BibitemShut
  {NoStop}%
\bibitem [{\citenamefont {H\"uttel}\ \emph {et~al.}(2009)\citenamefont
  {H\"uttel}, \citenamefont {Witkamp}, \citenamefont {Leijnse}, \citenamefont
  {Wegewijs},\ and\ \citenamefont {{van der Zant}}}]{Huettel2009}%
  \BibitemOpen
  \bibfield  {author} {\bibinfo {author} {\bibfnamefont {A.~K.}\ \bibnamefont
  {H\"uttel}}, \bibinfo {author} {\bibfnamefont {B.}~\bibnamefont {Witkamp}},
  \bibinfo {author} {\bibfnamefont {M.}~\bibnamefont {Leijnse}}, \bibinfo
  {author} {\bibfnamefont {M.~R.}\ \bibnamefont {Wegewijs}}, \ and\ \bibinfo
  {author} {\bibfnamefont {H.~S.~J.}\ \bibnamefont {{van der Zant}}},\
  }\href@noop {} {\bibfield  {journal} {\bibinfo  {journal} {Phys. Rev. Lett.}\
  }\textbf {\bibinfo {volume} {102}},\ \bibinfo {pages} {225501} (\bibinfo
  {year} {2009})}\BibitemShut {NoStop}%
\bibitem [{\citenamefont {Hihath}\ \emph {et~al.}(2010)\citenamefont {Hihath},
  \citenamefont {Bruot},\ and\ \citenamefont {Tao}}]{Tao2010}%
  \BibitemOpen
  \bibfield  {author} {\bibinfo {author} {\bibfnamefont {J.}~\bibnamefont
  {Hihath}}, \bibinfo {author} {\bibfnamefont {C.}~\bibnamefont {Bruot}}, \
  and\ \bibinfo {author} {\bibfnamefont {N.}~\bibnamefont {Tao}},\ }\href@noop
  {} {\bibfield  {journal} {\bibinfo  {journal} {ACS Nano}\ }\textbf {\bibinfo
  {volume} {4}},\ \bibinfo {pages} {3823} (\bibinfo {year} {2010})}\BibitemShut
  {NoStop}%
\bibitem [{\citenamefont {Ballmann}\ \emph {et~al.}(2010)\citenamefont
  {Ballmann}, \citenamefont {Hieringer}, \citenamefont {Secker}, \citenamefont
  {Zheng}, \citenamefont {Gladysz}, \citenamefont {G\"orling},\ and\
  \citenamefont {Weber}}]{Ballmann2010}%
  \BibitemOpen
  \bibfield  {author} {\bibinfo {author} {\bibfnamefont {S.}~\bibnamefont
  {Ballmann}}, \bibinfo {author} {\bibfnamefont {W.}~\bibnamefont {Hieringer}},
  \bibinfo {author} {\bibfnamefont {D.}~\bibnamefont {Secker}}, \bibinfo
  {author} {\bibfnamefont {Q.}~\bibnamefont {Zheng}}, \bibinfo {author}
  {\bibfnamefont {J.~A.}\ \bibnamefont {Gladysz}}, \bibinfo {author}
  {\bibfnamefont {A.}~\bibnamefont {G\"orling}}, \ and\ \bibinfo {author}
  {\bibfnamefont {H.~B.}\ \bibnamefont {Weber}},\ }\href@noop {} {\bibfield
  {journal} {\bibinfo  {journal} {Chem. Phys. Chem.}\ }\textbf {\bibinfo
  {volume} {11}},\ \bibinfo {pages} {2256} (\bibinfo {year}
  {2010})}\BibitemShut {NoStop}%
\bibitem [{\citenamefont {Jewell}\ \emph {et~al.}(2010)\citenamefont {Jewell},
  \citenamefont {Tierney}, \citenamefont {Baber}, \citenamefont {Iski},
  \citenamefont {Laha},\ and\ \citenamefont {Sykes}}]{Jewell2010}%
  \BibitemOpen
  \bibfield  {author} {\bibinfo {author} {\bibfnamefont {A.~D.}\ \bibnamefont
  {Jewell}}, \bibinfo {author} {\bibfnamefont {H.~L.}\ \bibnamefont {Tierney}},
  \bibinfo {author} {\bibfnamefont {A.~E.}\ \bibnamefont {Baber}}, \bibinfo
  {author} {\bibfnamefont {E.~V.}\ \bibnamefont {Iski}}, \bibinfo {author}
  {\bibfnamefont {M.~M.}\ \bibnamefont {Laha}}, \ and\ \bibinfo {author}
  {\bibfnamefont {E.~C.~H.}\ \bibnamefont {Sykes}},\ }\href@noop {} {\bibfield
  {journal} {\bibinfo  {journal} {J. Phys.: Condens. Matter}\ }\textbf
  {\bibinfo {volume} {22}},\ \bibinfo {pages} {264006} (\bibinfo {year}
  {2010})}\BibitemShut {NoStop}%
\bibitem [{\citenamefont {Osorio}\ \emph {et~al.}(2010)\citenamefont {Osorio},
  \citenamefont {Ruben}, \citenamefont {Seldenthuis}, \citenamefont {Lehn},\
  and\ \citenamefont {{van der Zant}}}]{Osorio2010}%
  \BibitemOpen
  \bibfield  {author} {\bibinfo {author} {\bibfnamefont {E.~A.}\ \bibnamefont
  {Osorio}}, \bibinfo {author} {\bibfnamefont {M.}~\bibnamefont {Ruben}},
  \bibinfo {author} {\bibfnamefont {J.~S.}\ \bibnamefont {Seldenthuis}},
  \bibinfo {author} {\bibfnamefont {J.~M.}\ \bibnamefont {Lehn}}, \ and\
  \bibinfo {author} {\bibfnamefont {H.~S.~J.}\ \bibnamefont {{van der Zant}}},\
  }\href@noop {} {\bibfield  {journal} {\bibinfo  {journal} {Small}\ }\textbf
  {\bibinfo {volume} {6}},\ \bibinfo {pages} {174} (\bibinfo {year}
  {2010})}\BibitemShut {NoStop}%
\bibitem [{\citenamefont {Koch}\ and\ \citenamefont {von
  Oppen}(2005)}]{Koch2005}%
  \BibitemOpen
  \bibfield  {author} {\bibinfo {author} {\bibfnamefont {J.}~\bibnamefont
  {Koch}}\ and\ \bibinfo {author} {\bibfnamefont {F.}~\bibnamefont {von
  Oppen}},\ }\href {\doibase 10.1103/PhysRevLett.94.206804} {\bibfield
  {journal} {\bibinfo  {journal} {Phys. Rev. Lett.}\ }\textbf {\bibinfo
  {volume} {94}},\ \bibinfo {pages} {206804} (\bibinfo {year}
  {2005})}\BibitemShut {NoStop}%
\bibitem [{\citenamefont {H\"artle}\ \emph {et~al.}(2009)\citenamefont
  {H\"artle}, \citenamefont {Benesch},\ and\ \citenamefont {Thoss}}]{Hartle09}%
  \BibitemOpen
  \bibfield  {author} {\bibinfo {author} {\bibfnamefont {R.}~\bibnamefont
  {H\"artle}}, \bibinfo {author} {\bibfnamefont {C.}~\bibnamefont {Benesch}}, \
  and\ \bibinfo {author} {\bibfnamefont {M.}~\bibnamefont {Thoss}},\
  }\href@noop {} {\bibfield  {journal} {\bibinfo  {journal} {Phys. Rev. Lett.}\
  }\textbf {\bibinfo {volume} {102}},\ \bibinfo {pages} {146801} (\bibinfo
  {year} {2009})}\BibitemShut {NoStop}%
\bibitem [{\citenamefont {Romano}\ \emph {et~al.}(2010)\citenamefont {Romano},
  \citenamefont {Gagliardi}, \citenamefont {Pecchia},\ and\ \citenamefont {{Di
  Carlo}}}]{Romano10}%
  \BibitemOpen
  \bibfield  {author} {\bibinfo {author} {\bibfnamefont {G.}~\bibnamefont
  {Romano}}, \bibinfo {author} {\bibfnamefont {A.}~\bibnamefont {Gagliardi}},
  \bibinfo {author} {\bibfnamefont {A.}~\bibnamefont {Pecchia}}, \ and\
  \bibinfo {author} {\bibfnamefont {A.}~\bibnamefont {{Di Carlo}}},\
  }\href@noop {} {\bibfield  {journal} {\bibinfo  {journal} {Phys. Rev. B}\
  }\textbf {\bibinfo {volume} {81}},\ \bibinfo {pages} {115438} (\bibinfo
  {year} {2010})}\BibitemShut {NoStop}%
\bibitem [{\citenamefont {Secker}\ \emph {et~al.}(2011)\citenamefont {Secker},
  \citenamefont {Wagner}, \citenamefont {Ballmann}, \citenamefont {H\"artle},
  \citenamefont {Thoss},\ and\ \citenamefont {Weber}}]{Secker2010}%
  \BibitemOpen
  \bibfield  {author} {\bibinfo {author} {\bibfnamefont {D.}~\bibnamefont
  {Secker}}, \bibinfo {author} {\bibfnamefont {S.}~\bibnamefont {Wagner}},
  \bibinfo {author} {\bibfnamefont {S.}~\bibnamefont {Ballmann}}, \bibinfo
  {author} {\bibfnamefont {R.}~\bibnamefont {H\"artle}}, \bibinfo {author}
  {\bibfnamefont {M.}~\bibnamefont {Thoss}}, \ and\ \bibinfo {author}
  {\bibfnamefont {H.~B.}\ \bibnamefont {Weber}},\ }\href@noop {} {\bibfield
  {journal} {\bibinfo  {journal} {Phys. Rev. Lett.}\ }\textbf {\bibinfo
  {volume} {106}},\ \bibinfo {pages} {136807} (\bibinfo {year}
  {2011})}\BibitemShut {NoStop}%
\bibitem [{\citenamefont {H\"artle}\ \emph
  {et~al.}(2011{\natexlab{a}})\citenamefont {H\"artle}, \citenamefont {Butzin},
  \citenamefont {Rubio-Pons},\ and\ \citenamefont {Thoss}}]{Hartle2011b}%
  \BibitemOpen
  \bibfield  {author} {\bibinfo {author} {\bibfnamefont {R.}~\bibnamefont
  {H\"artle}}, \bibinfo {author} {\bibfnamefont {M.}~\bibnamefont {Butzin}},
  \bibinfo {author} {\bibfnamefont {O.}~\bibnamefont {Rubio-Pons}}, \ and\
  \bibinfo {author} {\bibfnamefont {M.}~\bibnamefont {Thoss}},\ }\href@noop {}
  {\bibfield  {journal} {\bibinfo  {journal} {Phys. Rev. Lett.}\ }\textbf
  {\bibinfo {volume} {107}},\ \bibinfo {pages} {046802} (\bibinfo {year}
  {2011}{\natexlab{a}})}\BibitemShut {NoStop}%
\bibitem [{\citenamefont {Ballmann}\ \emph {et~al.}(2013)\citenamefont
  {Ballmann}, \citenamefont {Hieringer}, \citenamefont {H\"artle},
  \citenamefont {Coto}, \citenamefont {Bryce}, \citenamefont {G\"orling},
  \citenamefont {Thoss},\ and\ \citenamefont {Weber}}]{Ballmann2013b}%
  \BibitemOpen
  \bibfield  {author} {\bibinfo {author} {\bibfnamefont {S.}~\bibnamefont
  {Ballmann}}, \bibinfo {author} {\bibfnamefont {W.}~\bibnamefont {Hieringer}},
  \bibinfo {author} {\bibfnamefont {R.}~\bibnamefont {H\"artle}}, \bibinfo
  {author} {\bibfnamefont {P.~B.}\ \bibnamefont {Coto}}, \bibinfo {author}
  {\bibfnamefont {M.~R.}\ \bibnamefont {Bryce}}, \bibinfo {author}
  {\bibfnamefont {A.}~\bibnamefont {G\"orling}}, \bibinfo {author}
  {\bibfnamefont {M.}~\bibnamefont {Thoss}}, \ and\ \bibinfo {author}
  {\bibfnamefont {H.~B.}\ \bibnamefont {Weber}},\ }\href@noop {} {\bibfield
  {journal} {\bibinfo  {journal} {Phys. Status Solidi B}\ }\textbf {\bibinfo
  {volume} {250}},\ \bibinfo {pages} {2452} (\bibinfo {year}
  {2013})}\BibitemShut {NoStop}%
\bibitem [{\citenamefont {H\"artle}\ \emph
  {et~al.}(2013{\natexlab{a}})\citenamefont {H\"artle}, \citenamefont
  {Peskin},\ and\ \citenamefont {Thoss}}]{Hartle2013}%
  \BibitemOpen
  \bibfield  {author} {\bibinfo {author} {\bibfnamefont {R.}~\bibnamefont
  {H\"artle}}, \bibinfo {author} {\bibfnamefont {U.}~\bibnamefont {Peskin}}, \
  and\ \bibinfo {author} {\bibfnamefont {M.}~\bibnamefont {Thoss}},\
  }\href@noop {} {\bibfield  {journal} {\bibinfo  {journal} {Phys. Status
  Solidi B}\ }\textbf {\bibinfo {volume} {250}},\ \bibinfo {pages} {2365}
  (\bibinfo {year} {2013}{\natexlab{a}})}\BibitemShut {NoStop}%
\bibitem [{\citenamefont {Braig}\ and\ \citenamefont
  {Flensberg}(2003)}]{Braig2003}%
  \BibitemOpen
  \bibfield  {author} {\bibinfo {author} {\bibfnamefont {S.}~\bibnamefont
  {Braig}}\ and\ \bibinfo {author} {\bibfnamefont {K.}~\bibnamefont
  {Flensberg}},\ }\href {\doibase 10.1103/PhysRevB.68.205324} {\bibfield
  {journal} {\bibinfo  {journal} {Phys. Rev. B}\ }\textbf {\bibinfo {volume}
  {68}},\ \bibinfo {pages} {205324} (\bibinfo {year} {2003})}\BibitemShut
  {NoStop}%
\bibitem [{\citenamefont {Zazunov}\ \emph
  {et~al.}(2006{\natexlab{a}})\citenamefont {Zazunov}, \citenamefont
  {Feinberg},\ and\ \citenamefont {Martin}}]{Zazunov2006}%
  \BibitemOpen
  \bibfield  {author} {\bibinfo {author} {\bibfnamefont {A.}~\bibnamefont
  {Zazunov}}, \bibinfo {author} {\bibfnamefont {D.}~\bibnamefont {Feinberg}}, \
  and\ \bibinfo {author} {\bibfnamefont {T.}~\bibnamefont {Martin}},\
  }\href@noop {} {\bibfield  {journal} {\bibinfo  {journal} {Phys. Rev. B}\
  }\textbf {\bibinfo {volume} {73}},\ \bibinfo {pages} {115405} (\bibinfo
  {year} {2006}{\natexlab{a}})}\BibitemShut {NoStop}%
\bibitem [{\citenamefont {H\"artle}\ and\ \citenamefont
  {Thoss}(2011{\natexlab{a}})}]{Haertle11}%
  \BibitemOpen
  \bibfield  {author} {\bibinfo {author} {\bibfnamefont {R.}~\bibnamefont
  {H\"artle}}\ and\ \bibinfo {author} {\bibfnamefont {M.}~\bibnamefont
  {Thoss}},\ }\href@noop {} {\bibfield  {journal} {\bibinfo  {journal} {Phys.
  Rev. B}\ }\textbf {\bibinfo {volume} {83}},\ \bibinfo {pages} {115414}
  (\bibinfo {year} {2011}{\natexlab{a}})}\BibitemShut {NoStop}%
\bibitem [{\citenamefont {Erpenbeck}\ \emph {et~al.}(2015)\citenamefont
  {Erpenbeck}, \citenamefont {H\"artle},\ and\ \citenamefont
  {Thoss}}]{Erpenbeck2015}%
  \BibitemOpen
  \bibfield  {author} {\bibinfo {author} {\bibfnamefont {A.}~\bibnamefont
  {Erpenbeck}}, \bibinfo {author} {\bibfnamefont {R.}~\bibnamefont {H\"artle}},
  \ and\ \bibinfo {author} {\bibfnamefont {M.}~\bibnamefont {Thoss}},\ }\href
  {\doibase 10.1103/PhysRevB.91.195418} {\bibfield  {journal} {\bibinfo
  {journal} {Phys. Rev. B}\ }\textbf {\bibinfo {volume} {91}},\ \bibinfo
  {pages} {195418} (\bibinfo {year} {2015})}\BibitemShut {NoStop}%
\bibitem [{\citenamefont {H\"artle}\ \emph
  {et~al.}(2011{\natexlab{b}})\citenamefont {H\"artle}, \citenamefont {Butzin},
  \citenamefont {Rubio-Pons},\ and\ \citenamefont {Thoss}}]{Haertle11_3}%
  \BibitemOpen
  \bibfield  {author} {\bibinfo {author} {\bibfnamefont {R.}~\bibnamefont
  {H\"artle}}, \bibinfo {author} {\bibfnamefont {M.}~\bibnamefont {Butzin}},
  \bibinfo {author} {\bibfnamefont {O.}~\bibnamefont {Rubio-Pons}}, \ and\
  \bibinfo {author} {\bibfnamefont {M.}~\bibnamefont {Thoss}},\ }\href@noop {}
  {\bibfield  {journal} {\bibinfo  {journal} {Phys. Rev. Lett.}\ }\textbf
  {\bibinfo {volume} {107}},\ \bibinfo {pages} {046802} (\bibinfo {year}
  {2011}{\natexlab{b}})}\BibitemShut {NoStop}%
\bibitem [{\citenamefont {Ballmann}\ \emph {et~al.}(2012)\citenamefont
  {Ballmann}, \citenamefont {H\"artle}, \citenamefont {Coto}, \citenamefont
  {Elbing}, \citenamefont {Mayor}, \citenamefont {Bryce}, \citenamefont
  {Thoss},\ and\ \citenamefont {Weber}}]{Ballmann2012}%
  \BibitemOpen
  \bibfield  {author} {\bibinfo {author} {\bibfnamefont {S.}~\bibnamefont
  {Ballmann}}, \bibinfo {author} {\bibfnamefont {R.}~\bibnamefont {H\"artle}},
  \bibinfo {author} {\bibfnamefont {P.~B.}\ \bibnamefont {Coto}}, \bibinfo
  {author} {\bibfnamefont {M.}~\bibnamefont {Elbing}}, \bibinfo {author}
  {\bibfnamefont {M.}~\bibnamefont {Mayor}}, \bibinfo {author} {\bibfnamefont
  {M.~R.}\ \bibnamefont {Bryce}}, \bibinfo {author} {\bibfnamefont
  {M.}~\bibnamefont {Thoss}}, \ and\ \bibinfo {author} {\bibfnamefont {H.~B.}\
  \bibnamefont {Weber}},\ }\href@noop {} {\bibfield  {journal} {\bibinfo
  {journal} {Phys. Rev. Lett.}\ }\textbf {\bibinfo {volume} {109}},\ \bibinfo
  {pages} {056801} (\bibinfo {year} {2012})}\BibitemShut {NoStop}%
\bibitem [{\citenamefont {H\"artle}\ \emph
  {et~al.}(2013{\natexlab{b}})\citenamefont {H\"artle}, \citenamefont
  {Butzin},\ and\ \citenamefont {Thoss}}]{Haertle13}%
  \BibitemOpen
  \bibfield  {author} {\bibinfo {author} {\bibfnamefont {R.}~\bibnamefont
  {H\"artle}}, \bibinfo {author} {\bibfnamefont {M.}~\bibnamefont {Butzin}}, \
  and\ \bibinfo {author} {\bibfnamefont {M.}~\bibnamefont {Thoss}},\
  }\href@noop {} {\bibfield  {journal} {\bibinfo  {journal} {Phys. Rev. B}\
  }\textbf {\bibinfo {volume} {87}},\ \bibinfo {pages} {085422} (\bibinfo
  {year} {2013}{\natexlab{b}})}\BibitemShut {NoStop}%
\bibitem [{\citenamefont {Boese}\ and\ \citenamefont
  {Schoeller}(2001)}]{Schoeller01}%
  \BibitemOpen
  \bibfield  {author} {\bibinfo {author} {\bibfnamefont {D.}~\bibnamefont
  {Boese}}\ and\ \bibinfo {author} {\bibfnamefont {H.}~\bibnamefont
  {Schoeller}},\ }\href@noop {} {\bibfield  {journal} {\bibinfo  {journal}
  {Europhys. Lett.}\ }\textbf {\bibinfo {volume} {54}},\ \bibinfo {pages} {668}
  (\bibinfo {year} {2001})}\BibitemShut {NoStop}%
\bibitem [{\citenamefont {Koch}\ and\ \citenamefont {{von
  Oppen}}(2005)}]{Koch05b}%
  \BibitemOpen
  \bibfield  {author} {\bibinfo {author} {\bibfnamefont {J.}~\bibnamefont
  {Koch}}\ and\ \bibinfo {author} {\bibfnamefont {F.}~\bibnamefont {{von
  Oppen}}},\ }\href@noop {} {\bibfield  {journal} {\bibinfo  {journal} {Phys.
  Rev. B}\ }\textbf {\bibinfo {volume} {72}},\ \bibinfo {pages} {113308}
  (\bibinfo {year} {2005})}\BibitemShut {NoStop}%
\bibitem [{\citenamefont {Zazunov}\ \emph
  {et~al.}(2006{\natexlab{b}})\citenamefont {Zazunov}, \citenamefont
  {Feinberg},\ and\ \citenamefont {Martin}}]{Zazunov06}%
  \BibitemOpen
  \bibfield  {author} {\bibinfo {author} {\bibfnamefont {A.}~\bibnamefont
  {Zazunov}}, \bibinfo {author} {\bibfnamefont {D.}~\bibnamefont {Feinberg}}, \
  and\ \bibinfo {author} {\bibfnamefont {T.}~\bibnamefont {Martin}},\
  }\href@noop {} {\bibfield  {journal} {\bibinfo  {journal} {Phys. Rev. B}\
  }\textbf {\bibinfo {volume} {73}},\ \bibinfo {pages} {115405} (\bibinfo
  {year} {2006}{\natexlab{b}})}\BibitemShut {NoStop}%
\bibitem [{\citenamefont {Leijnse}\ and\ \citenamefont
  {Wegewijs}(2008)}]{Leijnse09}%
  \BibitemOpen
  \bibfield  {author} {\bibinfo {author} {\bibfnamefont {M.}~\bibnamefont
  {Leijnse}}\ and\ \bibinfo {author} {\bibfnamefont {M.~R.}\ \bibnamefont
  {Wegewijs}},\ }\href@noop {} {\bibfield  {journal} {\bibinfo  {journal}
  {Phys. Rev. B}\ }\textbf {\bibinfo {volume} {78}},\ \bibinfo {pages} {235424}
  (\bibinfo {year} {2008})}\BibitemShut {NoStop}%
\bibitem [{\citenamefont {H\"artle}\ and\ \citenamefont
  {Thoss}(2011{\natexlab{b}})}]{Hartle2010b}%
  \BibitemOpen
  \bibfield  {author} {\bibinfo {author} {\bibfnamefont {R.}~\bibnamefont
  {H\"artle}}\ and\ \bibinfo {author} {\bibfnamefont {M.}~\bibnamefont
  {Thoss}},\ }\href@noop {} {\bibfield  {journal} {\bibinfo  {journal} {Phys.
  Rev. B}\ }\textbf {\bibinfo {volume} {83}},\ \bibinfo {pages} {115414}
  (\bibinfo {year} {2011}{\natexlab{b}})}\BibitemShut {NoStop}%
\bibitem [{\citenamefont {Galperin}\ \emph {et~al.}(2005)\citenamefont
  {Galperin}, \citenamefont {Ratner},\ and\ \citenamefont
  {Nitzan}}]{Galperin2005}%
  \BibitemOpen
  \bibfield  {author} {\bibinfo {author} {\bibfnamefont {M.}~\bibnamefont
  {Galperin}}, \bibinfo {author} {\bibfnamefont {M.~A.}\ \bibnamefont
  {Ratner}}, \ and\ \bibinfo {author} {\bibfnamefont {A.}~\bibnamefont
  {Nitzan}},\ }\href@noop {} {\bibfield  {journal} {\bibinfo  {journal} {Nano
  Lett.}\ }\textbf {\bibinfo {volume} {5}},\ \bibinfo {pages} {125} (\bibinfo
  {year} {2005})}\BibitemShut {NoStop}%
\bibitem [{\citenamefont {Albrecht}\ \emph {et~al.}(2012)\citenamefont
  {Albrecht}, \citenamefont {Wang}, \citenamefont {M\"uhlbacher}, \citenamefont
  {Thoss},\ and\ \citenamefont {Komnik}}]{Thoss2012}%
  \BibitemOpen
  \bibfield  {author} {\bibinfo {author} {\bibfnamefont {K.~F.}\ \bibnamefont
  {Albrecht}}, \bibinfo {author} {\bibfnamefont {H.}~\bibnamefont {Wang}},
  \bibinfo {author} {\bibfnamefont {L.}~\bibnamefont {M\"uhlbacher}}, \bibinfo
  {author} {\bibfnamefont {M.}~\bibnamefont {Thoss}}, \ and\ \bibinfo {author}
  {\bibfnamefont {A.}~\bibnamefont {Komnik}},\ }\href@noop {} {\bibfield
  {journal} {\bibinfo  {journal} {Phys. Rev. B}\ }\textbf {\bibinfo {volume}
  {86}},\ \bibinfo {pages} {081412} (\bibinfo {year} {2012})}\BibitemShut
  {NoStop}%
\bibitem [{\citenamefont {Albrecht}\ \emph {et~al.}(2013)\citenamefont
  {Albrecht}, \citenamefont {Martin-Rodero}, \citenamefont {Monreal},
  \citenamefont {M\"uhlbacher},\ and\ \citenamefont
  {Levy~Yeyati}}]{Albrecht2013}%
  \BibitemOpen
  \bibfield  {author} {\bibinfo {author} {\bibfnamefont {K.~F.}\ \bibnamefont
  {Albrecht}}, \bibinfo {author} {\bibfnamefont {A.}~\bibnamefont
  {Martin-Rodero}}, \bibinfo {author} {\bibfnamefont {R.~C.}\ \bibnamefont
  {Monreal}}, \bibinfo {author} {\bibfnamefont {L.}~\bibnamefont
  {M\"uhlbacher}}, \ and\ \bibinfo {author} {\bibfnamefont {A.}~\bibnamefont
  {Levy~Yeyati}},\ }\href {\doibase 10.1103/PhysRevB.87.085127} {\bibfield
  {journal} {\bibinfo  {journal} {Phys. Rev. B}\ }\textbf {\bibinfo {volume}
  {87}},\ \bibinfo {pages} {085127} (\bibinfo {year} {2013})}\BibitemShut
  {NoStop}%
\bibitem [{\citenamefont {Wilner}\ \emph {et~al.}(2014)\citenamefont {Wilner},
  \citenamefont {Wang}, \citenamefont {Thoss},\ and\ \citenamefont
  {Rabani}}]{Wilner2014}%
  \BibitemOpen
  \bibfield  {author} {\bibinfo {author} {\bibfnamefont {E.~Y.}\ \bibnamefont
  {Wilner}}, \bibinfo {author} {\bibfnamefont {H.}~\bibnamefont {Wang}},
  \bibinfo {author} {\bibfnamefont {M.}~\bibnamefont {Thoss}}, \ and\ \bibinfo
  {author} {\bibfnamefont {E.}~\bibnamefont {Rabani}},\ }\href {\doibase
  10.1103/PhysRevB.89.205129} {\bibfield  {journal} {\bibinfo  {journal} {Phys.
  Rev. B}\ }\textbf {\bibinfo {volume} {89}},\ \bibinfo {pages} {205129}
  (\bibinfo {year} {2014})}\BibitemShut {NoStop}%
\bibitem [{\citenamefont {Cederbaum}\ and\ \citenamefont
  {Domcke}(1974)}]{Cederbaum74}%
  \BibitemOpen
  \bibfield  {author} {\bibinfo {author} {\bibfnamefont {L.~S.}\ \bibnamefont
  {Cederbaum}}\ and\ \bibinfo {author} {\bibfnamefont {W.}~\bibnamefont
  {Domcke}},\ }\href@noop {} {\bibfield  {journal} {\bibinfo  {journal} {J.
  Chem. Phys.}\ }\textbf {\bibinfo {volume} {60}},\ \bibinfo {pages} {2878}
  (\bibinfo {year} {1974})}\BibitemShut {NoStop}%
\bibitem [{\citenamefont {Benesch}\ \emph {et~al.}(2008)\citenamefont
  {Benesch}, \citenamefont {Cizek}, \citenamefont {Klimes}, \citenamefont
  {Thoss},\ and\ \citenamefont {Domcke}}]{Benesch08}%
  \BibitemOpen
  \bibfield  {author} {\bibinfo {author} {\bibfnamefont {C.}~\bibnamefont
  {Benesch}}, \bibinfo {author} {\bibfnamefont {M.}~\bibnamefont {Cizek}},
  \bibinfo {author} {\bibfnamefont {J.}~\bibnamefont {Klimes}}, \bibinfo
  {author} {\bibfnamefont {M.}~\bibnamefont {Thoss}}, \ and\ \bibinfo {author}
  {\bibfnamefont {W.}~\bibnamefont {Domcke}},\ }\href@noop {} {\bibfield
  {journal} {\bibinfo  {journal} {J. Phys. Chem. C}\ }\textbf {\bibinfo
  {volume} {112}},\ \bibinfo {pages} {9880} (\bibinfo {year}
  {2008})}\BibitemShut {NoStop}%
\bibitem [{\citenamefont {Han}(2010)}]{Han2010}%
  \BibitemOpen
  \bibfield  {author} {\bibinfo {author} {\bibfnamefont {J.~E.}\ \bibnamefont
  {Han}},\ }\href@noop {} {\bibfield  {journal} {\bibinfo  {journal} {Phys.
  Rev. B}\ }\textbf {\bibinfo {volume} {81}},\ \bibinfo {pages} {113106}
  (\bibinfo {year} {2010})}\BibitemShut {NoStop}%
\bibitem [{\citenamefont {Wegewijs}\ and\ \citenamefont
  {Nowack}(2005)}]{Wegewijs05}%
  \BibitemOpen
  \bibfield  {author} {\bibinfo {author} {\bibfnamefont {M.~R.}\ \bibnamefont
  {Wegewijs}}\ and\ \bibinfo {author} {\bibfnamefont {K.~C.}\ \bibnamefont
  {Nowack}},\ }\href@noop {} {\bibfield  {journal} {\bibinfo  {journal} {New J.
  Phys.}\ }\textbf {\bibinfo {volume} {7}},\ \bibinfo {pages} {239} (\bibinfo
  {year} {2005})}\BibitemShut {NoStop}%
\bibitem [{\citenamefont {White}\ and\ \citenamefont
  {Galperin}(2012)}]{White2012}%
  \BibitemOpen
  \bibfield  {author} {\bibinfo {author} {\bibfnamefont {A.~J.}\ \bibnamefont
  {White}}\ and\ \bibinfo {author} {\bibfnamefont {M.}~\bibnamefont
  {Galperin}},\ }\href@noop {} {\bibfield  {journal} {\bibinfo  {journal}
  {Phys. Chem. Chem. Phys.}\ }\textbf {\bibinfo {volume} {14}},\ \bibinfo
  {pages} {13809} (\bibinfo {year} {2012})}\BibitemShut {NoStop}%
\bibitem [{\citenamefont {Kaasbjerg}\ \emph {et~al.}(2013)\citenamefont
  {Kaasbjerg}, \citenamefont {Novotny},\ and\ \citenamefont
  {Nitzan}}]{Kaasbjerg2013}%
  \BibitemOpen
  \bibfield  {author} {\bibinfo {author} {\bibfnamefont {K.}~\bibnamefont
  {Kaasbjerg}}, \bibinfo {author} {\bibfnamefont {T.}~\bibnamefont {Novotny}},
  \ and\ \bibinfo {author} {\bibfnamefont {A.}~\bibnamefont {Nitzan}},\
  }\href@noop {} {\bibfield  {journal} {\bibinfo  {journal} {Phys. Rev. B}\
  }\textbf {\bibinfo {volume} {88}},\ \bibinfo {pages} {201405} (\bibinfo
  {year} {2013})}\BibitemShut {NoStop}%
\bibitem [{\citenamefont {Cizek}\ \emph {et~al.}(2005)\citenamefont {Cizek},
  \citenamefont {Thoss},\ and\ \citenamefont {Domcke}}]{Cizek05}%
  \BibitemOpen
  \bibfield  {author} {\bibinfo {author} {\bibfnamefont {M.}~\bibnamefont
  {Cizek}}, \bibinfo {author} {\bibfnamefont {M.}~\bibnamefont {Thoss}}, \ and\
  \bibinfo {author} {\bibfnamefont {W.}~\bibnamefont {Domcke}},\ }\href@noop {}
  {\bibfield  {journal} {\bibinfo  {journal} {Czech.\ J.\ Phys.}\ }\textbf
  {\bibinfo {volume} {55}},\ \bibinfo {pages} {189} (\bibinfo {year}
  {2005})}\BibitemShut {NoStop}%
\bibitem [{\citenamefont {Donarini}\ \emph {et~al.}(2006)\citenamefont
  {Donarini}, \citenamefont {Grifoni},\ and\ \citenamefont
  {Richter}}]{Donarini2006}%
  \BibitemOpen
  \bibfield  {author} {\bibinfo {author} {\bibfnamefont {A.}~\bibnamefont
  {Donarini}}, \bibinfo {author} {\bibfnamefont {M.}~\bibnamefont {Grifoni}}, \
  and\ \bibinfo {author} {\bibfnamefont {K.}~\bibnamefont {Richter}},\
  }\href@noop {} {\bibfield  {journal} {\bibinfo  {journal} {Phys. Rev. Lett.}\
  }\textbf {\bibinfo {volume} {97}},\ \bibinfo {pages} {166801} (\bibinfo
  {year} {2006})}\BibitemShut {NoStop}%
\bibitem [{\citenamefont {H\"ubener}\ and\ \citenamefont
  {Brandes}(2007)}]{Huebner2007}%
  \BibitemOpen
  \bibfield  {author} {\bibinfo {author} {\bibfnamefont {H.}~\bibnamefont
  {H\"ubener}}\ and\ \bibinfo {author} {\bibfnamefont {T.}~\bibnamefont
  {Brandes}},\ }\href@noop {} {\bibfield  {journal} {\bibinfo  {journal} {Phys.
  Rev. Lett.}\ }\textbf {\bibinfo {volume} {99}},\ \bibinfo {pages} {247206}
  (\bibinfo {year} {2007})}\BibitemShut {NoStop}%
\bibitem [{\citenamefont {Elste}\ and\ \citenamefont {von
  Oppen}(2008)}]{Elste2008}%
  \BibitemOpen
  \bibfield  {author} {\bibinfo {author} {\bibfnamefont {F.}~\bibnamefont
  {Elste}}\ and\ \bibinfo {author} {\bibfnamefont {F.}~\bibnamefont {von
  Oppen}},\ }\href {http://stacks.iop.org/1367-2630/10/i=6/a=065021} {\bibfield
   {journal} {\bibinfo  {journal} {New J. Phys.}\ }\textbf {\bibinfo {volume}
  {10}},\ \bibinfo {pages} {065021} (\bibinfo {year} {2008})}\BibitemShut
  {NoStop}%
\bibitem [{\citenamefont {Pshenichnyuk}\ and\ \citenamefont {\ifmmode
  \check{C}\else \v{C}\fi{}\'\i{}\ifmmode~\check{z}\else
  \v{z}\fi{}ek}(2011)}]{Pshenichnyuk2010}%
  \BibitemOpen
  \bibfield  {author} {\bibinfo {author} {\bibfnamefont {I.~A.}\ \bibnamefont
  {Pshenichnyuk}}\ and\ \bibinfo {author} {\bibfnamefont {M.}~\bibnamefont
  {\ifmmode \check{C}\else \v{C}\fi{}\'\i{}\ifmmode~\check{z}\else
  \v{z}\fi{}ek}},\ }\href@noop {} {\bibfield  {journal} {\bibinfo  {journal}
  {Phys. Rev. B}\ }\textbf {\bibinfo {volume} {83}},\ \bibinfo {pages} {165446}
  (\bibinfo {year} {2011})}\BibitemShut {NoStop}%
\bibitem [{\citenamefont {Br\"uggemann}\ \emph {et~al.}(2012)\citenamefont
  {Br\"uggemann}, \citenamefont {Weick}, \citenamefont {Pistolesi},\ and\
  \citenamefont {{von Oppen}}}]{Brueggemann2012}%
  \BibitemOpen
  \bibfield  {author} {\bibinfo {author} {\bibfnamefont {J.}~\bibnamefont
  {Br\"uggemann}}, \bibinfo {author} {\bibfnamefont {G.}~\bibnamefont {Weick}},
  \bibinfo {author} {\bibfnamefont {F.}~\bibnamefont {Pistolesi}}, \ and\
  \bibinfo {author} {\bibfnamefont {F.}~\bibnamefont {{von Oppen}}},\
  }\href@noop {} {\bibfield  {journal} {\bibinfo  {journal} {Phys. Rev. B}\
  }\textbf {\bibinfo {volume} {85}},\ \bibinfo {pages} {125441} (\bibinfo
  {year} {2012})}\BibitemShut {NoStop}%
\bibitem [{\citenamefont {Simine}\ and\ \citenamefont
  {Segal}(2014)}]{Simine2014}%
  \BibitemOpen
  \bibfield  {author} {\bibinfo {author} {\bibfnamefont {L.}~\bibnamefont
  {Simine}}\ and\ \bibinfo {author} {\bibfnamefont {D.}~\bibnamefont {Segal}},\
  }\href@noop {} {\bibfield  {journal} {\bibinfo  {journal} {J. Chem. Phys.}\
  }\textbf {\bibinfo {volume} {141}},\ \bibinfo {pages} {014704} (\bibinfo
  {year} {2014})}\BibitemShut {NoStop}%
\bibitem [{\citenamefont {Schmalz}\ \emph {et~al.}(2011)\citenamefont
  {Schmalz}, \citenamefont {Serrano-Andr\'es}, \citenamefont {Sauri},
  \citenamefont {Merch\'an},\ and\ \citenamefont {Oliva}}]{Schmalz2011}%
  \BibitemOpen
  \bibfield  {author} {\bibinfo {author} {\bibfnamefont {T.~G.}\ \bibnamefont
  {Schmalz}}, \bibinfo {author} {\bibfnamefont {L.}~\bibnamefont
  {Serrano-Andr\'es}}, \bibinfo {author} {\bibfnamefont {V.}~\bibnamefont
  {Sauri}}, \bibinfo {author} {\bibfnamefont {M.}~\bibnamefont {Merch\'an}}, \
  and\ \bibinfo {author} {\bibfnamefont {J.~M.}\ \bibnamefont {Oliva}},\
  }\href@noop {} {\bibfield  {journal} {\bibinfo  {journal} {J. Chem. Phys.}\
  }\textbf {\bibinfo {volume} {135}},\ \bibinfo {eid} {194103} (\bibinfo {year}
  {2011})}\BibitemShut {NoStop}%
\bibitem [{\citenamefont {Pozner}\ \emph {et~al.}(2014)\citenamefont {Pozner},
  \citenamefont {Lifshitz},\ and\ \citenamefont {Peskin}}]{Pozner2014}%
  \BibitemOpen
  \bibfield  {author} {\bibinfo {author} {\bibfnamefont {R.}~\bibnamefont
  {Pozner}}, \bibinfo {author} {\bibfnamefont {E.}~\bibnamefont {Lifshitz}}, \
  and\ \bibinfo {author} {\bibfnamefont {U.}~\bibnamefont {Peskin}},\
  }\href@noop {} {\bibfield  {journal} {\bibinfo  {journal} {Nano Lett.}\
  }\textbf {\bibinfo {volume} {14}},\ \bibinfo {pages} {6244} (\bibinfo {year}
  {2014})}\BibitemShut {NoStop}%
\bibitem [{\citenamefont {Cederbaum}\ and\ \citenamefont
  {Domcke}(1976)}]{Cederbaum76}%
  \BibitemOpen
  \bibfield  {author} {\bibinfo {author} {\bibfnamefont {L.~S.}\ \bibnamefont
  {Cederbaum}}\ and\ \bibinfo {author} {\bibfnamefont {W.}~\bibnamefont
  {Domcke}},\ }\href@noop {} {\bibfield  {journal} {\bibinfo  {journal} {J.
  Chem. Phys.}\ }\textbf {\bibinfo {volume} {64}},\ \bibinfo {pages} {603}
  (\bibinfo {year} {1976})}\BibitemShut {NoStop}%
\bibitem [{\citenamefont {Benesch}\ \emph {et~al.}(2006)\citenamefont
  {Benesch}, \citenamefont {Cizek}, \citenamefont {Thoss},\ and\ \citenamefont
  {Domcke}}]{Benesch06}%
  \BibitemOpen
  \bibfield  {author} {\bibinfo {author} {\bibfnamefont {C.}~\bibnamefont
  {Benesch}}, \bibinfo {author} {\bibfnamefont {M.}~\bibnamefont {Cizek}},
  \bibinfo {author} {\bibfnamefont {M.}~\bibnamefont {Thoss}}, \ and\ \bibinfo
  {author} {\bibfnamefont {W.}~\bibnamefont {Domcke}},\ }\href@noop {}
  {\bibfield  {journal} {\bibinfo  {journal} {Chem. Phys. Lett.}\ }\textbf
  {\bibinfo {volume} {430}},\ \bibinfo {pages} {355} (\bibinfo {year}
  {2006})}\BibitemShut {NoStop}%
\bibitem [{\citenamefont {Kondov}\ \emph {et~al.}(2007)\citenamefont {Kondov},
  \citenamefont {Cizek}, \citenamefont {Benesch}, \citenamefont {Thoss},\ and\
  \citenamefont {Wang}}]{Kondov07}%
  \BibitemOpen
  \bibfield  {author} {\bibinfo {author} {\bibfnamefont {I.}~\bibnamefont
  {Kondov}}, \bibinfo {author} {\bibfnamefont {M.}~\bibnamefont {Cizek}},
  \bibinfo {author} {\bibfnamefont {C.}~\bibnamefont {Benesch}}, \bibinfo
  {author} {\bibfnamefont {M.}~\bibnamefont {Thoss}}, \ and\ \bibinfo {author}
  {\bibfnamefont {H.}~\bibnamefont {Wang}},\ }\href@noop {} {\bibfield
  {journal} {\bibinfo  {journal} {J. Phys. Chem. C}\ }\textbf {\bibinfo
  {volume} {111}},\ \bibinfo {pages} {11970} (\bibinfo {year}
  {2007})}\BibitemShut {NoStop}%
\bibitem [{\citenamefont {Benesch}\ \emph {et~al.}(2009)\citenamefont
  {Benesch}, \citenamefont {Rode}, \citenamefont {Cizek}, \citenamefont
  {H\"artle}, \citenamefont {Rubio-Pons}, \citenamefont {Thoss},\ and\
  \citenamefont {Sobolewski}}]{Benesch2009}%
  \BibitemOpen
  \bibfield  {author} {\bibinfo {author} {\bibfnamefont {C.}~\bibnamefont
  {Benesch}}, \bibinfo {author} {\bibfnamefont {M.~F.}\ \bibnamefont {Rode}},
  \bibinfo {author} {\bibfnamefont {M.}~\bibnamefont {Cizek}}, \bibinfo
  {author} {\bibfnamefont {R.}~\bibnamefont {H\"artle}}, \bibinfo {author}
  {\bibfnamefont {O.}~\bibnamefont {Rubio-Pons}}, \bibinfo {author}
  {\bibfnamefont {M.}~\bibnamefont {Thoss}}, \ and\ \bibinfo {author}
  {\bibfnamefont {A.~L.}\ \bibnamefont {Sobolewski}},\ }\href@noop {}
  {\bibfield  {journal} {\bibinfo  {journal} {J. Phys. Chem. C}\ }\textbf
  {\bibinfo {volume} {113}},\ \bibinfo {pages} {10315} (\bibinfo {year}
  {2009})}\BibitemShut {NoStop}%
\bibitem [{\citenamefont {H\"artle}(2012)}]{HartlePhD}%
  \BibitemOpen
  \bibfield  {author} {\bibinfo {author} {\bibfnamefont {R.}~\bibnamefont
  {H\"artle}},\ }\emph {\bibinfo {title} {Vibrationally Coupled Electron
  Transport through Single-Molecule Junctions}},\ \href@noop {} {Ph.D.
  thesis},\ \bibinfo  {school} {Friedrich-Alexander-Universit\"at
  Erlangen-N\"urnberg} (\bibinfo {year} {2012})\BibitemShut {NoStop}%
\bibitem [{\citenamefont {Reckermann}\ \emph {et~al.}(2008)\citenamefont
  {Reckermann}, \citenamefont {Leijnse}, \citenamefont {Wegewijs},\ and\
  \citenamefont {Schoeller}}]{Reckermann2008}%
  \BibitemOpen
  \bibfield  {author} {\bibinfo {author} {\bibfnamefont {F.}~\bibnamefont
  {Reckermann}}, \bibinfo {author} {\bibfnamefont {M.}~\bibnamefont {Leijnse}},
  \bibinfo {author} {\bibfnamefont {M.~R.}\ \bibnamefont {Wegewijs}}, \ and\
  \bibinfo {author} {\bibfnamefont {H.}~\bibnamefont {Schoeller}},\ }\href@noop
  {} {\bibfield  {journal} {\bibinfo  {journal} {EPL}\ }\textbf {\bibinfo
  {volume} {83}},\ \bibinfo {pages} {58001} (\bibinfo {year}
  {2008})}\BibitemShut {NoStop}%
\bibitem [{\citenamefont {Frederiksen}\ \emph {et~al.}(2008)\citenamefont
  {Frederiksen}, \citenamefont {Franke}, \citenamefont {Arnau}, \citenamefont
  {Schulze}, \citenamefont {Pascual},\ and\ \citenamefont
  {Lorente}}]{Frederiksen2008}%
  \BibitemOpen
  \bibfield  {author} {\bibinfo {author} {\bibfnamefont {T.}~\bibnamefont
  {Frederiksen}}, \bibinfo {author} {\bibfnamefont {K.}~\bibnamefont {Franke}},
  \bibinfo {author} {\bibfnamefont {A.}~\bibnamefont {Arnau}}, \bibinfo
  {author} {\bibfnamefont {G.}~\bibnamefont {Schulze}}, \bibinfo {author}
  {\bibfnamefont {J.~I.}\ \bibnamefont {Pascual}}, \ and\ \bibinfo {author}
  {\bibfnamefont {N.}~\bibnamefont {Lorente}},\ }\href@noop {} {\bibfield
  {journal} {\bibinfo  {journal} {Phys. Rev. B}\ }\textbf {\bibinfo {volume}
  {78}},\ \bibinfo {pages} {233401} (\bibinfo {year} {2008})}\BibitemShut
  {NoStop}%
\bibitem [{\citenamefont {Schultz}\ \emph {et~al.}(2008)\citenamefont
  {Schultz}, \citenamefont {Nunner},\ and\ \citenamefont {von
  Oppen}}]{Schultz2008}%
  \BibitemOpen
  \bibfield  {author} {\bibinfo {author} {\bibfnamefont {M.~G.}\ \bibnamefont
  {Schultz}}, \bibinfo {author} {\bibfnamefont {T.~S.}\ \bibnamefont {Nunner}},
  \ and\ \bibinfo {author} {\bibfnamefont {F.}~\bibnamefont {von Oppen}},\
  }\href@noop {} {\bibfield  {journal} {\bibinfo  {journal} {Phys. Rev. B}\
  }\textbf {\bibinfo {volume} {77}},\ \bibinfo {pages} {075323} (\bibinfo
  {year} {2008})}\BibitemShut {NoStop}%
\bibitem [{\citenamefont {Repp}\ \emph {et~al.}(2010)\citenamefont {Repp},
  \citenamefont {Liljeroth},\ and\ \citenamefont {Meyer}}]{Repp2009}%
  \BibitemOpen
  \bibfield  {author} {\bibinfo {author} {\bibfnamefont {J.}~\bibnamefont
  {Repp}}, \bibinfo {author} {\bibfnamefont {P.}~\bibnamefont {Liljeroth}}, \
  and\ \bibinfo {author} {\bibfnamefont {G.}~\bibnamefont {Meyer}},\
  }\href@noop {} {\bibfield  {journal} {\bibinfo  {journal} {Nature Phys.}\
  }\textbf {\bibinfo {volume} {6}},\ \bibinfo {pages} {975} (\bibinfo {year}
  {2010})}\BibitemShut {NoStop}%
\bibitem [{\citenamefont {Cuevas}\ and\ \citenamefont
  {Scheer}(2010)}]{cuevasscheer2010}%
  \BibitemOpen
  \bibfield  {author} {\bibinfo {author} {\bibfnamefont {J.~C.}\ \bibnamefont
  {Cuevas}}\ and\ \bibinfo {author} {\bibfnamefont {E.}~\bibnamefont
  {Scheer}},\ }\href@noop {} {\emph {\bibinfo {title} {Molecular Electronics:
  An Introduction To Theory And Experiment}}}\ (\bibinfo  {publisher} {World
  Scientific},\ \bibinfo {address} {Singapore},\ \bibinfo {year}
  {2010})\BibitemShut {NoStop}%
\bibitem [{\citenamefont {\ifmmode \check{C}\else
  \v{C}\fi{}\'\i\ifmmode~\check{z}\else \v{z}\fi{}ek}\ \emph
  {et~al.}(2004)\citenamefont {\ifmmode \check{C}\else
  \v{C}\fi{}\'\i\ifmmode~\check{z}\else \v{z}\fi{}ek}, \citenamefont {Thoss},\
  and\ \citenamefont {Domcke}}]{Cizek2004}%
  \BibitemOpen
  \bibfield  {author} {\bibinfo {author} {\bibfnamefont {M.}~\bibnamefont
  {\ifmmode \check{C}\else \v{C}\fi{}\'\i\ifmmode~\check{z}\else
  \v{z}\fi{}ek}}, \bibinfo {author} {\bibfnamefont {M.}~\bibnamefont {Thoss}},
  \ and\ \bibinfo {author} {\bibfnamefont {W.}~\bibnamefont {Domcke}},\
  }\href@noop {} {\bibfield  {journal} {\bibinfo  {journal} {Phys. Rev. B}\
  }\textbf {\bibinfo {volume} {70}},\ \bibinfo {pages} {125406} (\bibinfo
  {year} {2004})}\BibitemShut {NoStop}%
\bibitem [{\citenamefont {Peskin}(2010)}]{Peskin2010}%
  \BibitemOpen
  \bibfield  {author} {\bibinfo {author} {\bibfnamefont {U.}~\bibnamefont
  {Peskin}},\ }\href@noop {} {\bibfield  {journal} {\bibinfo  {journal} {J.
  Phys. B}\ }\textbf {\bibinfo {volume} {43}},\ \bibinfo {pages} {153001}
  (\bibinfo {year} {2010})}\BibitemShut {NoStop}%
\bibitem [{\citenamefont {Mahan}(1981)}]{Mahan81}%
  \BibitemOpen
  \bibfield  {author} {\bibinfo {author} {\bibfnamefont {G.~D.}\ \bibnamefont
  {Mahan}},\ }\href@noop {} {\emph {\bibinfo {title} {Many-Particle Physics}}}\
  (\bibinfo  {publisher} {Plenum Press},\ \bibinfo {address} {New York},\
  \bibinfo {year} {1981})\BibitemShut {NoStop}%
\bibitem [{\citenamefont {Mitra}\ \emph {et~al.}(2004)\citenamefont {Mitra},
  \citenamefont {Aleiner},\ and\ \citenamefont {Millis}}]{Mitra04}%
  \BibitemOpen
  \bibfield  {author} {\bibinfo {author} {\bibfnamefont {A.}~\bibnamefont
  {Mitra}}, \bibinfo {author} {\bibfnamefont {I.}~\bibnamefont {Aleiner}}, \
  and\ \bibinfo {author} {\bibfnamefont {A.~J.}\ \bibnamefont {Millis}},\
  }\href@noop {} {\bibfield  {journal} {\bibinfo  {journal} {Phys. Rev. B}\
  }\textbf {\bibinfo {volume} {69}},\ \bibinfo {pages} {245302} (\bibinfo
  {year} {2004})}\BibitemShut {NoStop}%
\bibitem [{\citenamefont {Galperin}\ \emph {et~al.}(2006)\citenamefont
  {Galperin}, \citenamefont {Nitzan},\ and\ \citenamefont
  {Ratner}}]{Galperin06}%
  \BibitemOpen
  \bibfield  {author} {\bibinfo {author} {\bibfnamefont {M.}~\bibnamefont
  {Galperin}}, \bibinfo {author} {\bibfnamefont {A.}~\bibnamefont {Nitzan}}, \
  and\ \bibinfo {author} {\bibfnamefont {M.~A.}\ \bibnamefont {Ratner}},\
  }\href@noop {} {\bibfield  {journal} {\bibinfo  {journal} {Phys. Rev. B}\
  }\textbf {\bibinfo {volume} {73}},\ \bibinfo {pages} {045314} (\bibinfo
  {year} {2006})}\BibitemShut {NoStop}%
\bibitem [{\citenamefont {H\"artle}\ \emph {et~al.}(2008)\citenamefont
  {H\"artle}, \citenamefont {Benesch},\ and\ \citenamefont {Thoss}}]{Hartle}%
  \BibitemOpen
  \bibfield  {author} {\bibinfo {author} {\bibfnamefont {R.}~\bibnamefont
  {H\"artle}}, \bibinfo {author} {\bibfnamefont {C.}~\bibnamefont {Benesch}}, \
  and\ \bibinfo {author} {\bibfnamefont {M.}~\bibnamefont {Thoss}},\
  }\href@noop {} {\bibfield  {journal} {\bibinfo  {journal} {Phys. Rev. B}\
  }\textbf {\bibinfo {volume} {77}},\ \bibinfo {pages} {205314} (\bibinfo
  {year} {2008})}\BibitemShut {NoStop}%
\bibitem [{\citenamefont {May}(2002)}]{May02}%
  \BibitemOpen
  \bibfield  {author} {\bibinfo {author} {\bibfnamefont {V.}~\bibnamefont
  {May}},\ }\href@noop {} {\bibfield  {journal} {\bibinfo  {journal} {Phys.
  Rev. B}\ }\textbf {\bibinfo {volume} {66}},\ \bibinfo {pages} {245411}
  (\bibinfo {year} {2002})}\BibitemShut {NoStop}%
\bibitem [{\citenamefont {Lehmann}\ \emph {et~al.}(2004)\citenamefont
  {Lehmann}, \citenamefont {Kohler}, \citenamefont {May},\ and\ \citenamefont
  {{H\"anggi}}}]{Lehmann04}%
  \BibitemOpen
  \bibfield  {author} {\bibinfo {author} {\bibfnamefont {J.}~\bibnamefont
  {Lehmann}}, \bibinfo {author} {\bibfnamefont {S.}~\bibnamefont {Kohler}},
  \bibinfo {author} {\bibfnamefont {V.}~\bibnamefont {May}}, \ and\ \bibinfo
  {author} {\bibfnamefont {P.}~\bibnamefont {{H\"anggi}}},\ }\href@noop {}
  {\bibfield  {journal} {\bibinfo  {journal} {J. Chem. Phys.}\ }\textbf
  {\bibinfo {volume} {121}},\ \bibinfo {pages} {2278} (\bibinfo {year}
  {2004})}\BibitemShut {NoStop}%
\bibitem [{\citenamefont {Harbola}\ \emph {et~al.}(2006)\citenamefont
  {Harbola}, \citenamefont {Esposito},\ and\ \citenamefont
  {Mukamel}}]{Harbola2006}%
  \BibitemOpen
  \bibfield  {author} {\bibinfo {author} {\bibfnamefont {U.}~\bibnamefont
  {Harbola}}, \bibinfo {author} {\bibfnamefont {M.}~\bibnamefont {Esposito}}, \
  and\ \bibinfo {author} {\bibfnamefont {S.}~\bibnamefont {Mukamel}},\
  }\href@noop {} {\bibfield  {journal} {\bibinfo  {journal} {Phys. Rev. B}\
  }\textbf {\bibinfo {volume} {74}},\ \bibinfo {pages} {235309} (\bibinfo
  {year} {2006})}\BibitemShut {NoStop}%
\bibitem [{\citenamefont {Volkovich}\ \emph {et~al.}(2008)\citenamefont
  {Volkovich}, \citenamefont {{Caspary Toroker}},\ and\ \citenamefont
  {Peskin}}]{Volkovich2008}%
  \BibitemOpen
  \bibfield  {author} {\bibinfo {author} {\bibfnamefont {R.}~\bibnamefont
  {Volkovich}}, \bibinfo {author} {\bibfnamefont {M.}~\bibnamefont {{Caspary
  Toroker}}}, \ and\ \bibinfo {author} {\bibfnamefont {U.}~\bibnamefont
  {Peskin}},\ }\href@noop {} {\bibfield  {journal} {\bibinfo  {journal} {J.
  Chem. Phys.}\ }\textbf {\bibinfo {volume} {129}},\ \bibinfo {pages} {034501}
  (\bibinfo {year} {2008})}\BibitemShut {NoStop}%
\bibitem [{\citenamefont {H\"artle}\ \emph {et~al.}(2010)\citenamefont
  {H\"artle}, \citenamefont {Volkovich}, \citenamefont {Thoss},\ and\
  \citenamefont {Peskin}}]{Hartle2010}%
  \BibitemOpen
  \bibfield  {author} {\bibinfo {author} {\bibfnamefont {R.}~\bibnamefont
  {H\"artle}}, \bibinfo {author} {\bibfnamefont {R.}~\bibnamefont {Volkovich}},
  \bibinfo {author} {\bibfnamefont {M.}~\bibnamefont {Thoss}}, \ and\ \bibinfo
  {author} {\bibfnamefont {U.}~\bibnamefont {Peskin}},\ }\href@noop {}
  {\bibfield  {journal} {\bibinfo  {journal} {J. Chem. Phys.}\ }\textbf
  {\bibinfo {volume} {133}},\ \bibinfo {pages} {081102} (\bibinfo {year}
  {2010})}\BibitemShut {NoStop}%
\bibitem [{\citenamefont {Nakajima}(1958)}]{Nakajima}%
  \BibitemOpen
  \bibfield  {author} {\bibinfo {author} {\bibfnamefont {S.}~\bibnamefont
  {Nakajima}},\ }\href@noop {} {\bibfield  {journal} {\bibinfo  {journal}
  {Prog. Theor. Phys.}\ }\textbf {\bibinfo {volume} {20}},\ \bibinfo {pages}
  {948} (\bibinfo {year} {1958})}\BibitemShut {NoStop}%
\bibitem [{\citenamefont {Zwanzig}(1960)}]{Zwanzig}%
  \BibitemOpen
  \bibfield  {author} {\bibinfo {author} {\bibfnamefont {R.}~\bibnamefont
  {Zwanzig}},\ }\href@noop {} {\bibfield  {journal} {\bibinfo  {journal} {J.
  Chem. Phys.}\ }\textbf {\bibinfo {volume} {33}},\ \bibinfo {pages} {1338}
  (\bibinfo {year} {1960})}\BibitemShut {NoStop}%
\bibitem [{\citenamefont {H\"artle}\ and\ \citenamefont
  {Millis}(2014)}]{Hartle2014}%
  \BibitemOpen
  \bibfield  {author} {\bibinfo {author} {\bibfnamefont {R.}~\bibnamefont
  {H\"artle}}\ and\ \bibinfo {author} {\bibfnamefont {A.~J.}\ \bibnamefont
  {Millis}},\ }\href@noop {} {\bibfield  {journal} {\bibinfo  {journal} {Phys.
  Rev. B}\ }\textbf {\bibinfo {volume} {90}},\ \bibinfo {pages} {245426}
  (\bibinfo {year} {2014})}\BibitemShut {NoStop}%
\bibitem [{\citenamefont {Schinabeck}\ \emph {et~al.}(2014)\citenamefont
  {Schinabeck}, \citenamefont {H\"artle}, \citenamefont {Weber},\ and\
  \citenamefont {Thoss}}]{Schinabeck2014}%
  \BibitemOpen
  \bibfield  {author} {\bibinfo {author} {\bibfnamefont {C.}~\bibnamefont
  {Schinabeck}}, \bibinfo {author} {\bibfnamefont {R.}~\bibnamefont
  {H\"artle}}, \bibinfo {author} {\bibfnamefont {H.~B.}\ \bibnamefont {Weber}},
  \ and\ \bibinfo {author} {\bibfnamefont {M.}~\bibnamefont {Thoss}},\
  }\href@noop {} {\bibfield  {journal} {\bibinfo  {journal} {Phys. Rev. B}\
  }\textbf {\bibinfo {volume} {90}},\ \bibinfo {pages} {075409} (\bibinfo
  {year} {2014})}\BibitemShut {NoStop}%
\bibitem [{\citenamefont {Ryndyk}\ \emph {et~al.}(2006)\citenamefont {Ryndyk},
  \citenamefont {Hartung},\ and\ \citenamefont {Cuniberti}}]{Ryndyk06}%
  \BibitemOpen
  \bibfield  {author} {\bibinfo {author} {\bibfnamefont {D.~A.}\ \bibnamefont
  {Ryndyk}}, \bibinfo {author} {\bibfnamefont {M.}~\bibnamefont {Hartung}}, \
  and\ \bibinfo {author} {\bibfnamefont {G.}~\bibnamefont {Cuniberti}},\
  }\href@noop {} {\bibfield  {journal} {\bibinfo  {journal} {Phys. Rev. B}\
  }\textbf {\bibinfo {volume} {73}},\ \bibinfo {pages} {045420} (\bibinfo
  {year} {2006})}\BibitemShut {NoStop}%
\bibitem [{\citenamefont {Koch}\ \emph {et~al.}(2006)\citenamefont {Koch},
  \citenamefont {Semmelhack}, \citenamefont {{von Oppen}},\ and\ \citenamefont
  {Nitzan}}]{Semmelhack}%
  \BibitemOpen
  \bibfield  {author} {\bibinfo {author} {\bibfnamefont {J.}~\bibnamefont
  {Koch}}, \bibinfo {author} {\bibfnamefont {M.}~\bibnamefont {Semmelhack}},
  \bibinfo {author} {\bibfnamefont {F.}~\bibnamefont {{von Oppen}}}, \ and\
  \bibinfo {author} {\bibfnamefont {A.}~\bibnamefont {Nitzan}},\ }\href@noop {}
  {\bibfield  {journal} {\bibinfo  {journal} {Phys. Rev. B}\ }\textbf {\bibinfo
  {volume} {73}},\ \bibinfo {pages} {155306} (\bibinfo {year}
  {2006})}\BibitemShut {NoStop}%
\bibitem [{\citenamefont {Leturcq}\ \emph {et~al.}(2009)\citenamefont
  {Leturcq}, \citenamefont {Stampfer}, \citenamefont {Inderbitzin},
  \citenamefont {Durrer}, \citenamefont {Hierold}, \citenamefont {Mariani},
  \citenamefont {Schultz}, \citenamefont {von Oppen},\ and\ \citenamefont
  {Ensslin}}]{Leturcq2009}%
  \BibitemOpen
  \bibfield  {author} {\bibinfo {author} {\bibfnamefont {R.}~\bibnamefont
  {Leturcq}}, \bibinfo {author} {\bibfnamefont {C.}~\bibnamefont {Stampfer}},
  \bibinfo {author} {\bibfnamefont {K.}~\bibnamefont {Inderbitzin}}, \bibinfo
  {author} {\bibfnamefont {L.}~\bibnamefont {Durrer}}, \bibinfo {author}
  {\bibfnamefont {C.}~\bibnamefont {Hierold}}, \bibinfo {author} {\bibfnamefont
  {E.}~\bibnamefont {Mariani}}, \bibinfo {author} {\bibfnamefont {M.~G.}\
  \bibnamefont {Schultz}}, \bibinfo {author} {\bibfnamefont {F.}~\bibnamefont
  {von Oppen}}, \ and\ \bibinfo {author} {\bibfnamefont {K.}~\bibnamefont
  {Ensslin}},\ }\href@noop {} {\bibfield  {journal} {\bibinfo  {journal}
  {Nature Phys.}\ }\textbf {\bibinfo {volume} {5}},\ \bibinfo {pages} {327}
  (\bibinfo {year} {2009})}\BibitemShut {NoStop}%
\bibitem [{\citenamefont {H\"artle}\ and\ \citenamefont
  {Thoss}(2011{\natexlab{c}})}]{Hartle2011}%
  \BibitemOpen
  \bibfield  {author} {\bibinfo {author} {\bibfnamefont {R.}~\bibnamefont
  {H\"artle}}\ and\ \bibinfo {author} {\bibfnamefont {M.}~\bibnamefont
  {Thoss}},\ }\href@noop {} {\bibfield  {journal} {\bibinfo  {journal} {Phys.
  Rev. B}\ }\textbf {\bibinfo {volume} {83}},\ \bibinfo {pages} {125419}
  (\bibinfo {year} {2011}{\natexlab{c}})}\BibitemShut {NoStop}%
\bibitem [{\citenamefont {H\"artle}\ \emph
  {et~al.}(2013{\natexlab{c}})\citenamefont {H\"artle}, \citenamefont {Cohen},
  \citenamefont {Reichman},\ and\ \citenamefont {Millis}}]{Haertle2013}%
  \BibitemOpen
  \bibfield  {author} {\bibinfo {author} {\bibfnamefont {R.}~\bibnamefont
  {H\"artle}}, \bibinfo {author} {\bibfnamefont {G.}~\bibnamefont {Cohen}},
  \bibinfo {author} {\bibfnamefont {D.~R.}\ \bibnamefont {Reichman}}, \ and\
  \bibinfo {author} {\bibfnamefont {A.~J.}\ \bibnamefont {Millis}},\
  }\href@noop {} {\bibfield  {journal} {\bibinfo  {journal} {Phys. Rev. B}\
  }\textbf {\bibinfo {volume} {88}},\ \bibinfo {pages} {235426} (\bibinfo
  {year} {2013}{\natexlab{c}})}\BibitemShut {NoStop}%
\bibitem [{\citenamefont {Park}\ \emph {et~al.}(2000)\citenamefont {Park},
  \citenamefont {Park}, \citenamefont {Lim}, \citenamefont {Anderson},
  \citenamefont {Alivisatos},\ and\ \citenamefont {McEuen}}]{Park2000}%
  \BibitemOpen
  \bibfield  {author} {\bibinfo {author} {\bibfnamefont {H.}~\bibnamefont
  {Park}}, \bibinfo {author} {\bibfnamefont {J.}~\bibnamefont {Park}}, \bibinfo
  {author} {\bibfnamefont {A.~K.~L.}\ \bibnamefont {Lim}}, \bibinfo {author}
  {\bibfnamefont {E.~H.}\ \bibnamefont {Anderson}}, \bibinfo {author}
  {\bibfnamefont {A.~P.}\ \bibnamefont {Alivisatos}}, \ and\ \bibinfo {author}
  {\bibfnamefont {P.~L.}\ \bibnamefont {McEuen}},\ }\href@noop {} {\bibfield
  {journal} {\bibinfo  {journal} {Nature}\ }\textbf {\bibinfo {volume} {407}},\
  \bibinfo {pages} {57} (\bibinfo {year} {2000})}\BibitemShut {NoStop}%
\bibitem [{\citenamefont {Kouwenhoven}\ \emph {et~al.}(2001)\citenamefont
  {Kouwenhoven}, \citenamefont {Austing},\ and\ \citenamefont
  {Tarucha}}]{Kouwenhoven2001}%
  \BibitemOpen
  \bibfield  {author} {\bibinfo {author} {\bibfnamefont {L.~P.}\ \bibnamefont
  {Kouwenhoven}}, \bibinfo {author} {\bibfnamefont {D.~G.}\ \bibnamefont
  {Austing}}, \ and\ \bibinfo {author} {\bibfnamefont {S.}~\bibnamefont
  {Tarucha}},\ }\href {http://stacks.iop.org/0034-4885/64/i=6/a=201} {\bibfield
   {journal} {\bibinfo  {journal} {Rep. Prog. Phys.}\ }\textbf {\bibinfo
  {volume} {64}},\ \bibinfo {pages} {701} (\bibinfo {year} {2001})}\BibitemShut
  {NoStop}%
\bibitem [{\citenamefont {Sapmaz}\ \emph {et~al.}(2005)\citenamefont {Sapmaz},
  \citenamefont {Jarillo-Herrero}, \citenamefont {Blanter},\ and\ \citenamefont
  {{van der Zant}}}]{Sapmaz05}%
  \BibitemOpen
  \bibfield  {author} {\bibinfo {author} {\bibfnamefont {S.}~\bibnamefont
  {Sapmaz}}, \bibinfo {author} {\bibfnamefont {P.}~\bibnamefont
  {Jarillo-Herrero}}, \bibinfo {author} {\bibfnamefont {Y.~M.}\ \bibnamefont
  {Blanter}}, \ and\ \bibinfo {author} {\bibfnamefont {H.~S.~J.}\ \bibnamefont
  {{van der Zant}}},\ }\href@noop {} {\bibfield  {journal} {\bibinfo  {journal}
  {New J. Phys.}\ }\textbf {\bibinfo {volume} {7}},\ \bibinfo {pages} {243}
  (\bibinfo {year} {2005})}\BibitemShut {NoStop}%
\bibitem [{\citenamefont {Chae}\ \emph {et~al.}(2006)\citenamefont {Chae},
  \citenamefont {Berry}, \citenamefont {Jung}, \citenamefont {Cotton},
  \citenamefont {Murillo},\ and\ \citenamefont {Yao}}]{Chae2006}%
  \BibitemOpen
  \bibfield  {author} {\bibinfo {author} {\bibfnamefont {D.-H.}\ \bibnamefont
  {Chae}}, \bibinfo {author} {\bibfnamefont {J.~F.}\ \bibnamefont {Berry}},
  \bibinfo {author} {\bibfnamefont {S.}~\bibnamefont {Jung}}, \bibinfo {author}
  {\bibfnamefont {F.~A.}\ \bibnamefont {Cotton}}, \bibinfo {author}
  {\bibfnamefont {C.~A.}\ \bibnamefont {Murillo}}, \ and\ \bibinfo {author}
  {\bibfnamefont {Z.}~\bibnamefont {Yao}},\ }\href@noop {} {\bibfield
  {journal} {\bibinfo  {journal} {Nano Lett.}\ }\textbf {\bibinfo {volume}
  {6}},\ \bibinfo {pages} {165} (\bibinfo {year} {2006})}\BibitemShut {NoStop}%
\bibitem [{\citenamefont {Thijssen}\ and\ \citenamefont {Van~der
  Zant}(2008)}]{Thijssen2008}%
  \BibitemOpen
  \bibfield  {author} {\bibinfo {author} {\bibfnamefont {J.~M.}\ \bibnamefont
  {Thijssen}}\ and\ \bibinfo {author} {\bibfnamefont {H.~S.~J.}\ \bibnamefont
  {Van~der Zant}},\ }\href {\doibase 10.1002/pssb.200743470} {\bibfield
  {journal} {\bibinfo  {journal} {Phys. Status Solidi B}\ }\textbf {\bibinfo
  {volume} {245}},\ \bibinfo {pages} {1455} (\bibinfo {year}
  {2008})}\BibitemShut {NoStop}%
\bibitem [{\citenamefont {Ballmann}\ and\ \citenamefont
  {Weber}(2012)}]{Ballmann2012_2}%
  \BibitemOpen
  \bibfield  {author} {\bibinfo {author} {\bibfnamefont {S.}~\bibnamefont
  {Ballmann}}\ and\ \bibinfo {author} {\bibfnamefont {H.~B.}\ \bibnamefont
  {Weber}},\ }\href@noop {} {\bibfield  {journal} {\bibinfo  {journal} {New J.
  Phys.}\ }\textbf {\bibinfo {volume} {14}},\ \bibinfo {pages} {123028}
  (\bibinfo {year} {2012})}\BibitemShut {NoStop}%
\end{thebibliography}%

\end{document}